\documentclass[aps,pra,reprint,superscriptaddress,nofootinbib]{revtex4-2}

\usepackage{amsmath,amssymb,mathtools,bm}
\usepackage{graphicx}
\usepackage{booktabs}
\usepackage{xcolor}
\usepackage[colorlinks=true,citecolor=blue,linkcolor=magenta]{hyperref}

\usepackage{microtype}

\usepackage{amsthm}
\usepackage{amsmath}
\usepackage[caption=false]{subfig}
\newtheorem{corollary}{Corollary}

\newtheorem{proposition}{Proposition}

\hypersetup{colorlinks=true,citecolor=blue!55!black,linkcolor=blue!55!black,urlcolor=blue!55!black}
\newcommand{\ket}[1]{\left|#1\right\rangle}
\newcommand{\bra}[1]{\left\langle#1\right|}
\newcommand{\proj}[1]{\ket{#1}\!\bra{#1}}

\begin{document}

\title{Optimal four-copy detection of the three-tangle and its fault-tolerant cost}

\author{Federico Tom\'as B. P\'erez}
\affiliation{Information Sciences, Los Alamos National Laboratory, Los Alamos, NM 87545, USA}
\affiliation{IFLP - CONICET, Departamento de F\'isica, Facultad de Ciencias Exactas,
Universidad Nacional de La Plata, C.C. 67, 1900 La Plata, Argentina}

\author{Berta Casas}
\affiliation{Information Sciences, Los Alamos National Laboratory, Los Alamos, NM 87545, USA}

\author{I-Chi Chen}
\affiliation{Information Sciences, Los Alamos National Laboratory, Los Alamos, NM 87545, USA}

\author{Su Yeon Chang}
\affiliation{Theoretical Division, Los Alamos National Laboratory, Los Alamos, NM 87545, USA}

\author{Philipp Schleich}
\affiliation{Theoretical Division, Los Alamos National Laboratory, Los Alamos, NM 87545, USA}

\author{M. Cerezo}
\thanks{cerezo@lanl.gov}
\affiliation{Information Sciences, Los Alamos National Laboratory, Los Alamos, NM 87545, USA}
\affiliation{Quantum Science Center, Oak Ridge, TN 37931, USA}

\begin{abstract}
The three-tangle, $\tau_3$, is a measure of genuine tripartite entanglement that distinguishes GHZ-type from W-type entanglement in pure three-qubit states.
Here, we identify the statistically optimal one-sided four-copy detector of the three-tangle and study its implementation cost. 
The optimal detector has click probability $\tau_3^2/48$, while previous four-copy constructions achieve $\tau_3^2/256$ and therefore require asymptotically $16/3$ times as many experimental repetitions.
Using the representation-theoretic structure of the problem, we construct an ancilla-free implementation of the optimal detector using 15 CNOT gates, together with an 11-CNOT detector with click probability $\tau_3^2/96$. 
These results reveal a tradeoff between statistical efficiency and implementation cost, since reducing the number of repetitions requires increasingly expensive circuits. 
We then show that this tradeoff persists once the circuits are encoded. 
In a fully fault-tolerant setting, the increased click probability comes with a substantial non-Clifford synthesis cost, while an early fault-tolerant implementation based on $T$ gates and STAR rotations reduces this overhead at the expense of additional encoded resources.
We quantify these costs using both the Steane and $[[8,3,2]]$ codes, providing explicit resource estimates for implementing the optimal detector across different computational regimes.
\end{abstract}

% \begin{abstract}

% The three-tangle, $\tau_3$, is a measure of genuine tripartite entanglement that distinguishes GHZ-type from W-type entanglement in pure three-qubit states. Here we identify the statistically optimal one-sided four-copy detector of the three-tangle and study the cost of implementing it across different quantum-computing regimes. The optimal detector has click probability $\tau_3^2/48$, while previous four-copy constructions achieve $\tau_3^2/256$ and therefore can asymptotically require $16/3$ times as many experimental repetitions. From the representation-theoretic structure of the problem, we construct an ancilla-free implementation of the optimal detector which uses 15 CNOT gates with CNOT depth three, together with an 11-CNOT detector with click probability $\tau_3^2/96$. We then consider the cost of implementing those circuits beyond the bare circuit model. In a fault-tolerant implementation, previous constructions can be implemented solely via Clifford gates, whereas the optimal detector necessarily requires non-Clifford resources. Here, we quantify the rotation-synthesis precision required to retain its statistical advantage over the fully Clifford circuit. We further consider its implementation in an early fault-tolerant setting using encoded logical qubits and partially protected rotations. Taken together, our results uncover a tradeoff between statistical efficiency and implementation cost.
% \end{abstract}

\maketitle

\section{Introduction}

Quantum information in theory can be quite different from quantum information in practice. For instance, once a density matrix is known, quantities such as purities, polynomial invariants, and entanglement measures often can be simply evaluated by a formula. However, applying this methodology in an experiment requires reconstructing the full state, which in turn can be far more demanding than evaluating the quantity of interest itself.  
For many quantities of interest, this tomographical reconstruction is unnecessary. 
Instead, they can be accessed directly through collective measurements on several copies of the same state. Some examples include direct measurements of nonlinear functions of a density matrix~\cite{ekert2002direct,brun2004measuring}, polynomial invariants~\cite{leifer2004measuring}, two-copy measurements of concurrence and multipartite concurrence~\cite{aolita2006measuring,walborn2006experimental}, and tests for product structure and pure-state entanglement~\cite{harrow2013testing,oszmaniec2013universal,beckey2021computable,schatzki2022hierarchy,beckey2023multipartite}. 

\begin{figure*}
    \centering
\includegraphics[width=1\linewidth]{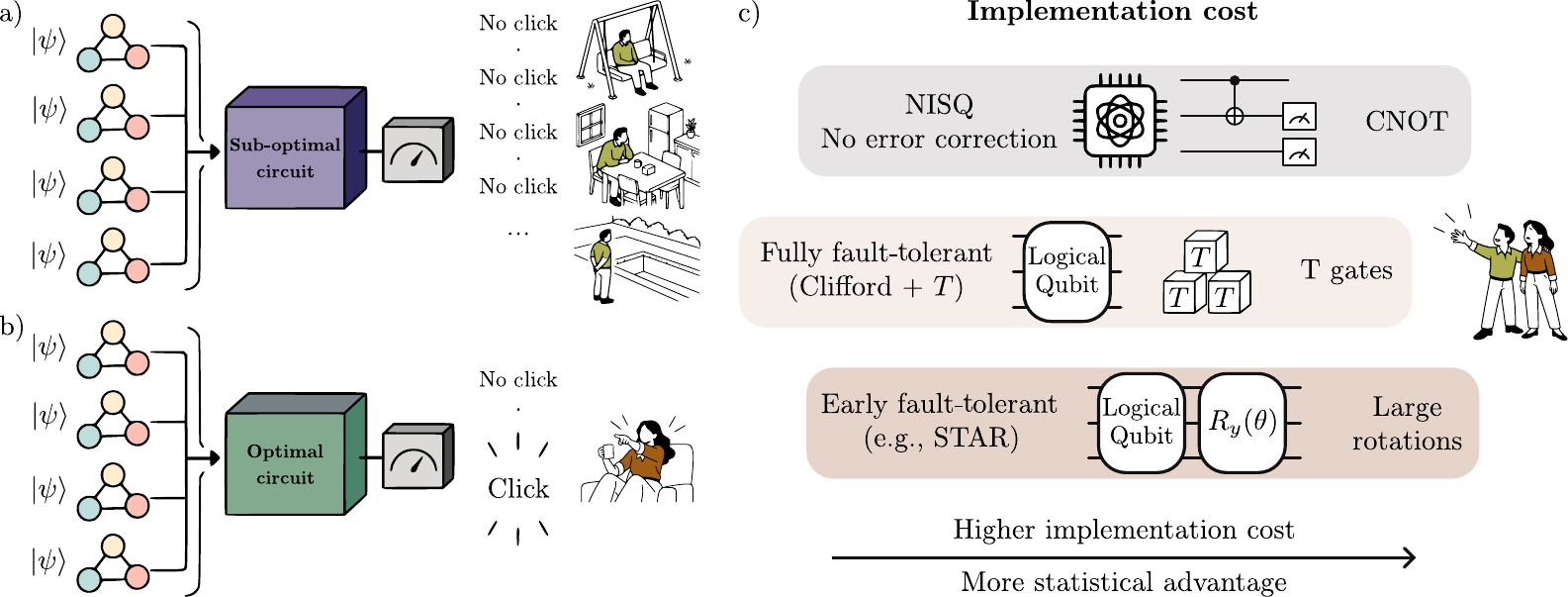}
    \caption{\textbf{Schematic representation of our results.} Four copies of the same three-qubit state are processed by a collective measurement designed to detect a nonzero three-tangle. (a) A suboptimal detector has a smaller click probability, and can therefore require many repeated circuit executions before a successful detection event is observed. (b) The optimal one-sided four-copy detector maximizes the click probability for a fixed $\tau_3$, substantially reducing the number of repetitions required to observe a click. (c) The resulting tradeoff between statistical efficiency and implementation cost persists across different computational regimes. On NISQ hardware, a larger click probability comes at the cost of additional entangling gates. In a fully fault-tolerant implementation, the same statistical advantage requires additional non-Clifford resources, while in an early fault-tolerant implementation these can be reduced using partially protected rotations at the cost of additional encoded resources. Thus, reducing the number of repetitions generally comes at an increased cost per circuit.}  \label{fig:schematic}
\end{figure*}

Once a direct circuit measurement is available for a given quantity of interest, a natural second question arises. \textit{How large is the resulting detection probability?} Different circuits can target the same quantity while producing the relevant outcome with very different probabilities, which directly affects the number of state preparations and circuit repetitions required for its estimation (see Fig.~\ref{fig:schematic}). This motivates going beyond the existence of a direct measurement and asking for circuits that are optimal in sample complexity. Importantly, however, the circuit with the largest detection probability need not be the cheapest one to implement. Reducing the number of experimental repetitions can instead come at the cost of additional gates and, once the circuit is encoded, additional fault-tolerant resources, posing a tradeof between statistical optimality and implementation costs.

Here we tackle the search for sample-optimal circuits for the three-tangle of pure three-qubit states~\cite{coffman2000distributed}, where we recall that the three-tangle distinguishes the GHZ entanglement class from the W class and its closure~\cite{dur2000three,miyake2003classification}. Several direct approaches to its measurement are known~\cite{yu2007measurable,yu2008measurable,perez2020measuring,yahyavi2022quantum,yu2016measurable,liu2026entanglement}, and some have even been deployed on cloud-based quantum computers~\cite{elyasi2025implementing}. In particular, Refs.~\cite{yu2007measurable,yu2008measurable} showed that the three-tangle can be obtained from local projective measurements on four copies via an ancilla-free circuit that uses six CNOT gates. The gate count was improved to five CNOT gates in the four-copy, also ancilla-free, circuit of~\cite{yahyavi2022quantum}.

More specifically, we seek to determine the four-copy measurement that uses the fewest repetitions among one-sided detectors. That is, detectors that never click on a pure state with zero three-tangle. Our first result is to use tools from pure-state property testing to identify the optimal perfect-completeness measurement. For the zero-three-tangle family, this leaves a single direction of four-copy rejection. We determine its normalization and find an optimal click probability $\tau_3^2/48$. We then show that this is larger by a factor of $16/3$ than the useful event probability in the constructions of~\cite{yu2007measurable,yahyavi2022quantum}, indicating that the latter are sub-optimal. Finally, we give two explicit circuits for estimating the three-tangle. The first realizes the optimal detector with 15~CNOT gates, while the second uses 11~CNOT gates and retains half of the optimal signal. Together, these results show that reducing the number of repetitions can come at the cost of a more demanding circuit implementation.

Finally, we consider how this tradeoff changes once the circuits are implemented fault-tolerantly. In the standard Clifford+$T$ model, the previous constructions in~\cite{yu2007measurable,yahyavi2022quantum}remain entirely Clifford, while our circuits require the synthesis of non-Clifford rotations. We quantify the corresponding synthesis cost and determine the precision required so that our detectors retain their statistical advantage. We also consider an early fault-tolerant implementation based on a combination of fault-tolerant $T$ gates and STAR rotations, and give explicit code-level resource estimates using the $[[7,1,3]]$ Steane code and $[[8,3,2]]$ code.

\section{Three-qubit entanglement}

In quantum resource theory of entanglement~\cite{chitambar2019quantum,diaz2025unified}, one seeks to understand the transformations between quantum states that are possible using local operations and classical communication (LOCC)~\cite{horodecki2009quantum,chitambar2014everything}. In such a LOCC protocol, each spatially separated party can perform local quantum operations, communicate experiment outcomes with each other and use this information to determine subsequent operations. While the mathematical characterization of LOCC can become quite complex~\cite{chitambar2014everything}, the situation becomes rather transparent if one focuses on a single branch of a LOCC protocol. For instance, in the three party case, where we dub the local experimenters Alice, Bob and Charlie, two pure states are equivalent under stochastic local operations and classical communication (SLOCC) if they can be converted into one another with nonzero probability~\cite{dur2000three}. That is, given $\ket{\psi},\ket{\phi}\in\mathcal{H}$, with $\mathcal{H}=(\mathbb{C}^{ 2})^{\otimes 3}$, they are SLOCC equivalent if
\begin{equation}
    \ket{\phi} \propto A\otimes B\otimes C\ket{\psi},
    \qquad
    A,B,C\in GL(2,\mathbb{C}).
\end{equation}
Notably, for three qubits, there are two classes of genuine tripartite entanglement that are inequivalent under this relation, which represented by the GHZ and W states~\cite{dur2000three},
\begin{equation}
    \ket{\mathrm{GHZ}} = \frac{\ket{000}+\ket{111}}{\sqrt{2}},
    \qquad
    \ket{W} = \frac{\ket{001}+\ket{010}+\ket{100}}{\sqrt{3}}.
    \nonumber
\end{equation}

In this context, the three-tangle $\tau_3$ was introduced as the residual three-party entanglement that remains after subtracting the pairwise contributions from the entanglement between one qubit and the other two~\cite{coffman2000distributed}. For a general pure three-qubit state
\begin{equation}
    \ket{\psi} = \sum_{i,j,k=0}^{1} a_{ijk}\ket{ijk},
    \label{eq:state_psi}
\end{equation}
the three-tangle is given by
\begin{equation}
    \tau_3(\psi)=4|\Delta(\psi)|,
\end{equation}
where $\Delta$ is Cayley's hyperdeterminant~\cite{coffman2000distributed,miyake2003classification},
\begin{align}
\Delta(\psi)
={}&
a_{000}^{2}a_{111}^{2}
+a_{001}^{2}a_{110}^{2}
\nonumber\\
&+a_{010}^{2}a_{101}^{2}
+a_{100}^{2}a_{011}^{2}
\nonumber\\
&-2a_{000}a_{001}a_{110}a_{111}
-2a_{000}a_{010}a_{101}a_{111}
\nonumber\\
&-2a_{000}a_{100}a_{011}a_{111}
-2a_{001}a_{010}a_{101}a_{110}
\nonumber\\
&-2a_{001}a_{100}a_{011}a_{110}
-2a_{010}a_{100}a_{011}a_{101}
\nonumber\\
&+4a_{000}a_{011}a_{101}a_{110}
+4a_{001}a_{010}a_{100}a_{111}.
\end{align}
As we can see, the hyperdeterminant is a homogeneous polynomial of degree four in the state amplitudes. Moreover, under SLOCC transformations it obeys
\begin{equation}\label{eq:transform}
    \Delta\!\left(
    A\otimes B\otimes C\ket{\psi}
    \right)
    =
    (\det A)^2(\det B)^2(\det C)^2
    \Delta(\psi).
\end{equation}
Hence, $\Delta$ is invariant under $SL(2,\mathbb{C})^{\otimes 3}$, while its vanishing is preserved under the full SLOCC group $GL(2,\mathbb{C})^{\otimes 3}$. In particular,
\begin{equation}
    \tau_3(\psi)>0
\end{equation}
for states in the GHZ SLOCC class, whereas
\begin{equation}
    \tau_3(\psi)=0
\end{equation}
for the W class and for the biseparable and fully separable classes~\cite{dur2000three,miyake2003classification}. The three-tangle therefore provides a natural way of distinguishing GHZ-type entanglement from the other pure three-qubit entanglement classes.

\section{The optimal four-copy detector}
Having introduced the three-tangle, we now ask how well it can be detected directly from four copies of the state. We consider a two-outcome measurement $\{E,\mathbb{I}-E\}$ acting on
$\ket{\psi}^{\otimes 4}$, where a click of $E$ is interpreted as detecting a nonzero three-tangle. We require the detector to be one-sided, so that
\begin{equation}
    \bra{\psi}^{\otimes 4} E \ket{\psi}^{\otimes 4} = 0
\end{equation}
for every pure three-qubit state $\ket{\psi}$ satisfying $\tau_3(\psi)=0$. Moreover, among all instruments satisfying this condition, we seek the one that maximizes the click probability for states with nonzero three-tangle.

The strategy for building the experiment starts by noting that the hyperdeterminant is a homogeneous polynomial of degree four in the amplitudes of $\ket{\psi}$. Then, the Riesz representation theorem~\cite{rudin1991functional}  implies that there exists a symmetric four-copy vector $\ket{w_\Delta}\in\mathcal{H}^{\otimes 4}$ such that
\begin{equation}
    \Delta(\psi) = \langle w_\Delta \ket{\psi}^{\otimes 4}.
    \label{eq:delta-four-copy}
\end{equation}
The vector $\ket{w_\Delta}$ can be readily obtained by identifying each quartic monomial in $\Delta$ with the corresponding symmetrized four-copy basis state. Explicitly, let $n_r(i,j,k,l)$ denote the number of times the label $r$ appears in the tuple $(i,j,k,l)$, where $i,j,k,l\in\{1,\ldots,8\}$ label the computational basis of the three-qubit Hilbert space. We define the normalized symmetric four-copy state as
\begin{equation}
    \ket{S(i,j,k,l)} = \sqrt{\frac{4!}{\prod_{r=1}^{8} n_r!}}\, P_{\rm sym}^{(4)} \ket{i,j,k,l},
\end{equation}
where
\begin{equation}
    P_{\rm sym}^{(4)} = \frac{1}{4!}\sum_{\pi\in S_4} U_\pi
\end{equation}
is the projector onto the symmetric subspace of the four copies and $U_\pi$ is a circuit that implements the permutation $\pi$ of the copies.  Using the convention $\ket{1}\equiv\ket{000},\cdots, \ket{8}\equiv\ket{111}$, we obtain
\begin{align}
    \ket{w_\Delta}
    =
    \frac{1}{\sqrt{6}}\Big(
    &
     \ket{S(1,1,8,8)}
    +\ket{S(2,2,7,7)}
    \nonumber\\
    &
    +\ket{S(3,3,6,6)}
    +\ket{S(4,4,5,5)}
    \nonumber\\
    &
    -\ket{S(1,2,7,8)}
    -\ket{S(1,3,6,8)}
    \nonumber\\
    &
    -\ket{S(1,4,5,8)}
    -\ket{S(2,3,6,7)}
    \nonumber\\
    &
    -\ket{S(2,4,5,7)}
    -\ket{S(3,4,5,6)}
    \nonumber\\
    &
    +2\ket{S(1,4,6,7)}
    +2\ket{S(2,3,5,8)}
    \Big)\,.
    \label{eq:wDelta-explicit}
\end{align}
Since the symmetric states appearing in Eq.~\eqref{eq:wDelta-explicit} are mutually orthogonal, its norm follows directly as
\begin{equation}
    \langle w_\Delta|w_\Delta\rangle =3.
    \label{eq:wdelta-norm}
\end{equation}
Then, as proved in Appendix~\ref{app:proof-prop-1}, the following proposition holds.

\begin{proposition}[Optimal four-copy detector]\label{prop:optimallity}
Let $E$ be any four-copy effect satisfying $0\leq E\leq\mathbb{I}$ and 
\begin{equation}
    \bra{\phi}^{\otimes 4} E \ket{\phi}^{\otimes 4} = 0
\end{equation}
for every pure three-qubit state with $\tau_3(\phi)=0$. Then
\begin{equation}
    P_{\mathrm{sym}}^{(4)} E P_{\mathrm{sym}}^{(4)}
    \leq Q_4,
    \label{eq:optimal-effect-bound}
\end{equation}
where
\begin{equation}
    Q_4 = \frac{\ket{w_\Delta}\!\bra{w_\Delta}}{3}.
    \label{eq:Q4-optimal}
\end{equation}
Consequently, for every pure three-qubit state,
\begin{equation}
    \bra{\psi}^{\otimes 4} E \ket{\psi}^{\otimes 4} \leq
    \frac{\tau_3(\psi)^2}{48}.
    \label{eq:optimal-probability-bound}
\end{equation}
The bound is saturated by $E=Q_4$.
\end{proposition}

\begin{figure}
    \centering
\includegraphics[width=1\columnwidth]{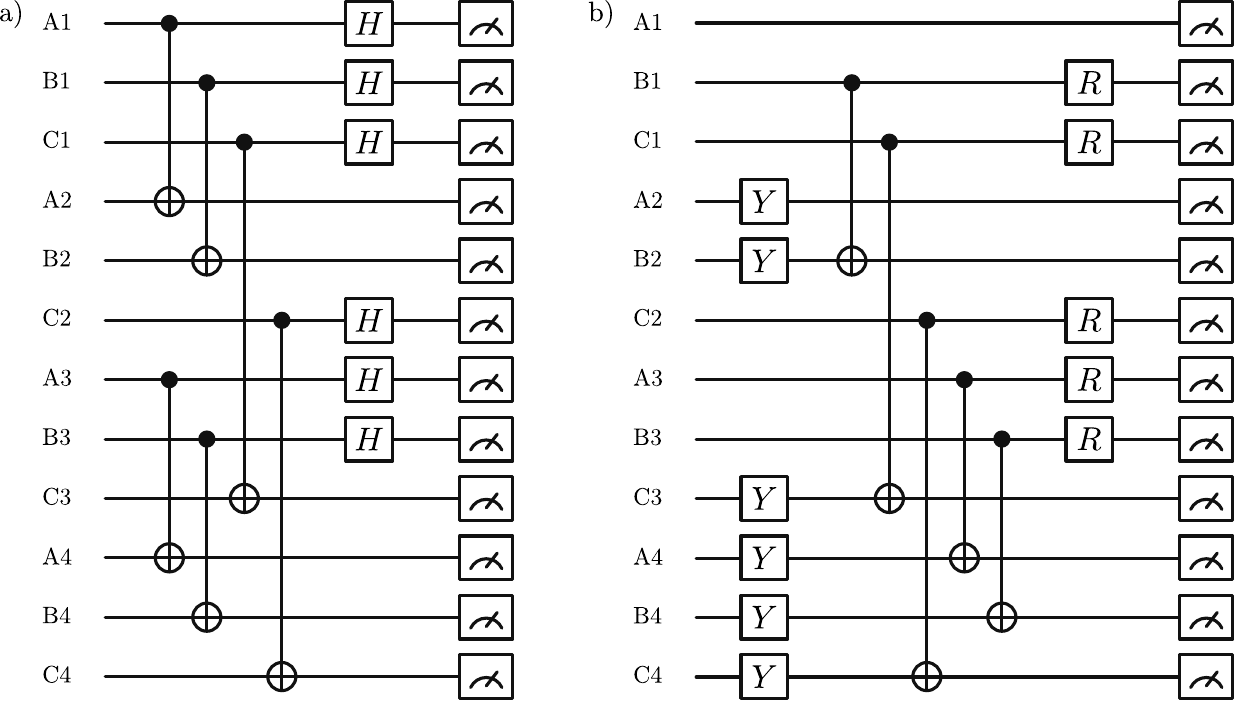}
    \caption{
\textbf{Previous ancilla-free four-copy circuits for the direct measurement of the three-tangle.}
(a) Circuit representation of the construction in~\cite{yu2007measurable}, which uses standard destructive Bell-basis measurements and translates into 6 CNOTS. 
(b) Circuit proposed by in~\cite{yahyavi2022quantum}, containing 5 CNOTs. Here, the $R$ gate denotes the rotation about the $y$ axis $R=R_y(-\pi/2)$. Throughout this work we use the convention $R_Y(\theta)=e^{-i\theta Y/2}$. Moreover, $A1$, $A2$, $A3$ and $A4$ denote the four copies of the first qubit in $\ket{\psi}$ (i.e., the Alice qubits), with similar notation for the copies of Bob and Charlie's qubits.}
\label{fig:previous-four-copy}
\end{figure}

\begin{figure*}
    \centering
\includegraphics[width=\textwidth]{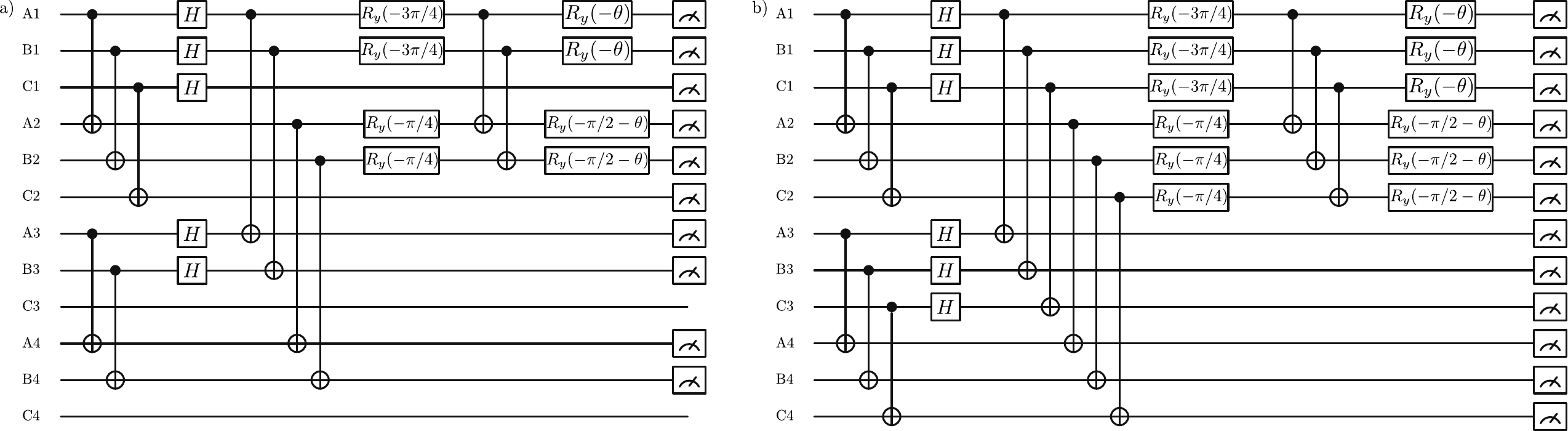}
    \caption{
    \textbf{Four-copy ancilla-free circuits introduced in this work.}
    (a) Reduced detector using two five-CNOT local $J=0$ projectors and one single-CNOT singlet measurement, for a total of $11$ CNOT gates and click probability $\tau_3^2/96$.
    (b) Optimal detector using three five-CNOT local $J=0$ projectors, for a total of $15$ CNOT gates and click probability $\tau_3^2/48$.
    Here $\theta=\arctan(1/\sqrt{2})$.
    }
    \label{fig:new-circuits}
\end{figure*}

Proposition~\ref{prop:optimallity} provides an upper bound on the click probability of any one-sided four-copy detector. 
In particular, the optimal probability depends only on the three-tangle and is
\begin{equation}
    q_{\mathrm{opt}}(\psi) = \frac{\tau_3(\psi)^2}{48}.
    \label{eq:qopt-final}
\end{equation}
For comparison, previous four-copy constructions for the direct measurement of the three-tangle~\cite{yu2007measurable,yahyavi2022quantum} are shown in Fig.~\ref{fig:previous-four-copy}. As proved in Appendix~\ref{app:prior}, these constructions yield a total click probability
\begin{equation}
    q_{\mathrm{prev}}(\psi)
    =
    \frac{\tau_3(\psi)^2}{256}.
\end{equation}
They, therefore, access the same quantity, but use only a fraction
\begin{equation}
    \frac{q_{\mathrm{prev}}}{q_{\mathrm{opt}}}
    =
    \frac{3}{16}
\end{equation}
of the probability available to an optimal one-sided four-copy detector. 

The difference in success probability has a direct experimental meaning. 
Independent repetition of $N$ four-copy trials, from an experiment with associated probability $q$, gives at least one click with probability $1-(1-q)^N$.
To obtain confidence $1-\delta$, when the state has three-tangle $\tau_3(\psi)$, one needs
\begin{equation}
N =\left\lceil
\frac{\ln\delta}{\ln(1-q)}
\right\rceil.
\label{eq:repetitions}
\end{equation}
For small $\tau_3(\psi)$, this scales as $\ln(1/\delta)/q$, but the difference is sizable even at $\tau_3(\psi)=1$.
For instance, at $95\%$ confidence, the ideal click statistics require 143 trials for the optimal detector and 766 trials for a detector with $q_{\mathrm{prev}}$.

\section{Four-copy circuit implementations}

We now provide two ancilla-free circuits for measuring the three-tangle, depicted in Fig.~\ref{fig:new-circuits}. The first circuit reduces the number of entangling gates at the cost of a smaller click probability, while the second realizes the optimal detector $Q_4$. In particular, the following proposition holds, which is proved in Appendix~\ref{app:proof-prop-2}.

\begin{proposition}[Four-copy circuit implementations]\label{prop:circuits}
The circuits in Fig.~\ref{fig:new-circuits} define one-sided four-copy detectors for the three-tangle. The circuit in Fig.~\ref{fig:new-circuits}(a) uses $11$ CNOT gates and has click probability
\begin{equation}
    q_{11}(\psi)
    =
    \frac{\tau_3(\psi)^2}{96}.
    \label{eq:q11}
\end{equation}
The circuit in Fig.~\ref{fig:new-circuits}(b) uses $15$ CNOT gates and realizes the optimal detector $Q_4$, with probability
\begin{equation}
    q_{15}(\psi)
    =
    q_{\mathrm{opt}}(\psi)
    =
    \frac{\tau_3(\psi)^2}{48}.
    \label{eq:q15}
\end{equation}
\end{proposition}

As further detailed in Appendix~\ref{app:rep-theory}, our circuits follow from the local total-spin-zero ($J=0$) sector of the four copies held by each party. From a representation-theoretic perspective, the construction amounts to projecting onto the trivial irreduciple representation contained in the relevant four-copy sector. For the optimal detector, the four qubits belonging to each $A$, $B$, and $C$ are measured with a five-CNOT projector onto this sector. The three projectors can be executed in parallel, resulting in overall $15$ CNOT gates. The reduced circuit replaces one of these projectors by a single-CNOT singlet measurement, lowering the CNOT gate count to $11$ while retaining one half of the optimal click probability. 

The resulting tradeoff is summarized in Table~\ref{tab:four-copy-comparison}. The $11$-CNOT construction requires twice as many repetitions as the optimal detector in the small-signal regime, but still substantially increases the click probability relative to the previous fixed four-copy constructions~\cite{yu2007measurable,yahyavi2022quantum}.
\begin{table*}[t]
\caption{
Fixed four-copy detectors discussed in the text.
The quantity $N/N_{\rm opt}$ gives the asymptotic repetition overhead
(see Eq.~\eqref{eq:repetitions}) relative to the optimal detector in the
small-signal regime. In the table FTQC denotes fault-tolerant quantum computation. 
For the fully FTQC implementation, the quoted $T$-counts for
our circuits assume $\tau_3=1$ and use the synthesis precision required
to retain a click probability at least as large as the Clifford
construction of Ref.~\cite{yu2007measurable}.
We note that while Ref.~\cite{liu2026entanglement} is not ancilla-free,
we have added it for comparison. In the early fault tolerant resource estimation, we also consider the repeat until success protocol that we implement STAR $R_z\left ( 2\theta \right )$ rotation gates if the STAR injection $R_z\left ( \theta \right )$ fails. The number shown in this column means the average number of $T$ gate and STAR rotation gates. 
}
\label{tab:four-copy-comparison}
\begin{ruledtabular}
\begin{tabular}{lcccccc}
Circuit
&
$q(\psi)$
&
CNOTs
&
$N/N_{\rm opt}$
&
Early FTQC
&
FTQC $T$ gates  & FTQC Cliffords
\\
\midrule % \hline
Ref.~\cite{liu2026entanglement}$^*$
&
$\tau_3^2/2048$
&
$44$
&
$128/3$
&
--
&
--
&
--
\\
Ref.~\cite{yu2007measurable}
&
$\tau_3^2/256$
&
$6$
&
$16/3$
&
0
&
$0$
&
$12$
\\
Ref.~\cite{yahyavi2022quantum}
&
$\tau_3^2/256$
&
$5$
&
$16/3$
&
0
&
$0$
&
$16$
\\
Fig.~\ref{fig:new-circuits}(a)
&
$\tau_3^2/96$
&
$11$
&
$2$
&
9.1T+21.7 STAR
&
$\simeq 127$
&
$\simeq150$
\\
Fig.~\ref{fig:new-circuits}(b)
&
$\tau_3^2/48$
&
$15$
&
$1$
&
13.7T+32.6 STAR
&
$\simeq 177$
&
$\simeq210$
\end{tabular}
\end{ruledtabular}
\label{tab: Ftqc_cost}
\end{table*}

\section{Fault-tolerant implementation}

In the previous sections we have discussed both the statistical efficiency and the CNOT cost of different circuits for measuring the three-tangle. At the level of bare circuits, the fixed four-copy constructions in previous works~\cite{yu2007measurable,yahyavi2022quantum} use five or six CNOT gates and have click probability $\tau_3^2/256$. Increasing the number of entangling gates to 11 CNOTs raises the click probability to $\tau_3^2/96$, while the 15-CNOT circuit reaches the optimal value $\tau_3^2/48$. Which point in this tradeoff is preferable will ultimately depend on the experimental setting. On noisy hardware, for instance, reducing the number of entangling gates can outweigh the cost of additional repetitions.

How does this tradeoff change once the circuits are encoded? We first consider a fully fault-tolerant implementation in the standard Clifford+$T$ model and determine the non-Clifford cost required to retain the statistical advantage of our circuits. We then consider an early fault-tolerant implementation based on STAR rotations. Finally, we give explicit code-level implementations and estimate their physical-qubit requirements.

\subsection{Fully fault-tolerant implementation}
\label{sec:fully-ft}
First, we look at a fully fault-tolerant implementation in the standard Clifford+$T$ model. For this analysis we adopt an idealized code-level description in which logical Clifford operations, syndrome extraction, and error correction are assumed to be available fault tolerantly, and we do not attempt to account for their code-dependent physical overhead. Instead, we focus on the additional non-Clifford resources required by each circuit, using the number of $T$ gates as our cost metric. Indeed, in many fault-tolerant architectures, logical Clifford operations can be implemented efficiently, while non-Clifford operations require additional resources such as magic-state preparation and distillation~\cite{bravyi2005universal,bravyi2012magic,fowler2012surface}. To determine resource requirements in this paradigm, gates must be compiled into the Clifford+$T$ gate set. Exact Clifford+$T$ synthesis is possible only for a restricted set of unitaries~\cite{giles2013exact,casas2025matchgate}, while arbitrary rotations need to be approximated up to some prescribed precision~\cite{ross2014optimal,bosco2026quantum}.

For the circuits considered here, this distinction is particularly simple. The constructions of Refs.~\cite{yu2007measurable,yahyavi2022quantum} contain only Clifford operations. By contrast, both of our proposed circuits shown in Fig.~\ref{fig:new-circuits} contain the angle
\begin{equation}
    \theta=\arctan\left(\frac{1}{\sqrt{2}}\right),
\end{equation}
which cannot be implemented exactly by a Clifford+$T$ circuit of finite length. Still, we can leverage standard algorithms, such as \textsc{gridsynth}~\cite{ross2014optimal}, to estimate the number of $T$ gates required to approximate this gate with error at most $\epsilon$ in operator norm. We find that the following proposition holds.

\begin{proposition}[Clifford+$T$ implementation cost]
\label{prop:clifford-t-cost}
The four-copy circuits of Refs.~\cite{yu2007measurable,yahyavi2022quantum} require no $T$ gates. For the two circuits introduced in this work, approximating each occurrence of the arbitrary rotation $R_y(\theta)$ with operator-norm error at most $\epsilon$ using \textsc{gridsynth}~\cite{ross2014optimal} gives the asymptotic $T$-count scalings
\begin{align}
    \#_T^{(11)} &= 12\log_2(1/\epsilon)+\mathcal{O}\!\left(\log\log(1/\epsilon)\right),\\
    \#_T^{(15)} &= 18\log_2(1/\epsilon)+\mathcal{O}\!\left(\log\log(1/\epsilon)\right),
\end{align}
for the 11-CNOT and optimal 15-CNOT detectors, respectively.
\end{proposition}

The proof of this result is given in Appendix~\ref{app:fault-tolerant}. Proposition~\ref{prop:clifford-t-cost} shows that the tradeoff takes a rather different form in fully fault-tolerant quantum computation. The constructions of Refs.~\cite{yu2007measurable,yahyavi2022quantum} remain entirely Clifford, whereas the increase in click probability provided by our circuits now comes with a non-Clifford synthesis cost that grows logarithmically with the required precision.

Interestingly, this non-Clifford cost is not simply a consequence of the particular decomposition used in Fig.~\ref{fig:new-circuits}. An exact finite Clifford+$T$ implementation of the optimal detector is impossible under the assumptions considered here. To see this, consider the GHZ state with $\tau_3=1$. The optimal detector clicks with probability
\begin{equation}
    q_{\rm opt}(\mathrm{GHZ})=\frac{1}{48}.
\end{equation}
The state $\ket{\mathrm{GHZ}}^{\otimes4}$ is a stabilizer state. For any finite Clifford+$T$ circuit acting on a stabilizer input, the resulting amplitudes belong to the ring $\mathbb{Z}[1/\sqrt{2},i]$~\cite{giles2013exact}, and computational-basis event probabilities therefore belong to $\mathbb{Z}[1/2,\sqrt{2}]$. Whenever such a probability is rational, its denominator in lowest terms must be a power of two. Since $1/48$ is not dyadic, no finite exact Clifford+$T$ circuit with stabilizer ancillas, computational-basis readout, and no postselection can realize the optimal detector. Additional details can be found in Appendix~\ref{app:fault-tolerant}.

The relevant question is then not whether the optimal detector can be synthesized exactly, but how accurately it needs to be synthesized before its larger click probability compensates for its additional fault-tolerant cost. To answer this question we can take as reference point the Clifford construction of Ref.~\cite{yu2007measurable}. The following corollary gives exactly the precision required to retain the statistical advantage.

\begin{corollary}[Precision required so that optimal circuits retain advantage over Clifford-only circuits]
\label{cor:ft-precision}
Suppose that every occurrence of the arbitrary rotation $R_y(\pm\theta)$ is synthesized with operator-norm error at most $\epsilon$. A sufficient condition for the 11-CNOT circuit to retain a click probability at least as large as that of Ref.~\cite{yu2007measurable} is
\begin{equation}
    \epsilon\leq\frac{5}{6144}\tau_3(\psi)^2.
    \label{eq:precision-11}
\end{equation}
For the optimal 15-CNOT circuit, a sufficient condition is
\begin{equation}
    \epsilon\leq\frac{13}{9216}\tau_3(\psi)^2.
    \label{eq:precision-15}
\end{equation}
\end{corollary}

The proof of this result is given in Appendix~\ref{app:fault-tolerant}. For $\tau_3=1$, Eqs.~\eqref{eq:precision-11} and~\eqref{eq:precision-15} give
\begin{equation}
    \epsilon_{11}\lesssim8.1\times10^{-4},
    \qquad
    \epsilon_{15}\lesssim1.4\times10^{-3}.
\end{equation}
Using the typical \textsc{gridsynth} scaling of Proposition~\ref{prop:clifford-t-cost}, and retaining the four and six exact $T$ gates appearing in the respective circuits, these precisions correspond to approximately
\begin{equation}
    \#_{T}^{(11)}\simeq127,
    \qquad
    \#_{T}^{(15)}\simeq177.
\end{equation}
Thus, the rotations do not need to be synthesized to extremely high precision simply to retain a statistical advantage over the Clifford construction. Moreover, denoting as $\#_{C}$ the number of Cliffords needed for the same implementation, we find
\begin{equation}
    \#_{C}^{(11)}\simeq150,
    \qquad
    \#_{C}^{(15)}\simeq210\,,
\end{equation}
with the same counts being equal to $12$ and $16$ for the circuits in~\cite{yu2007measurable} and~\cite{yahyavi2022quantum}, respectively. These results are presented in Table~\ref{tab:four-copy-comparison}. Additional details are provided in the Appendix.

\subsection{Early fault-tolerant implementation}

The Clifford+$T$ implementation above provides a fully fault-tolerant realization of our circuits, but the synthesis of the arbitrary rotations leads to a comparatively large $T$-gate cost. We now consider whether these rotations can be implemented more cheaply in an early fault-tolerant setting. To this end, we use the space-time efficient analog rotation (STAR) approach~\cite{toshio2025practical,ismail2026transversal}, which implements small-angle logical rotations using partially protected resource states. As in the fully fault-tolerant construction, the $R_y$ rotations appearing in Fig.~\ref{fig:new-circuits} can be converted to $R_z$ rotations by Clifford conjugation, so it is sufficient to consider the synthesis of $R_z(\theta)$.

Directly implementing the relatively large rotations appearing in our circuits using STAR alone can incur substantial logical errors. Following the hybrid approach of Ref.~\cite{chen2026efficient}, we instead combine fault-tolerant $T$ gates with smaller-angle rotations implemented using STAR. Specifically, we approximate the target rotation through a truncated signed-binary expansion,
\begin{equation}
     R_Z(\theta)\approx\prod_{k=2}^{K}\left[R_Z\left(\frac{\pi}{2^k}\right)\right]^{C_k},
\end{equation}
where $C_k\in\{-1,0,1\}$ and the coefficients are chosen such that
\begin{equation}
\left\|R_Z(\theta)-\prod_{k=2}^{K}\left[R_Z\left(\frac{\pi}{2^k}\right)\right]^{C_k}\right\|\leq\epsilon_{\mathrm{syn}}.
\end{equation}
Here, $\epsilon_{\mathrm{syn}}$ denotes the synthesis error. Under the convention $R_Z(\theta)=e^{-i\theta Z/2}$, the $k=2$ factor corresponds to a $T$ or $T^\dagger$ gate up to a global phase, while the remaining nonzero terms are implemented using STAR. For the rotation angle $\theta=-\arctan(1/\sqrt{2})$, the STAR+$T$ synthesis yields
\begin{equation}
    R_Z(\theta)\approx R_Z\left(-\frac{\pi}{4}\right)R_Z\left(\frac{\pi}{16}\right)R_Z\left(-\frac{\pi}{128}\right)R_Z\left(-\frac{\pi}{2048}\right).\nonumber
\end{equation}

The angular mismatch in this approximation is approximately $3.53\times10^{-4}$, corresponding to an operator-norm error of approximately $1.77\times10^{-4}$. This is below the sufficient precision bounds in Eqs.~\eqref{eq:precision-11} and~\eqref{eq:precision-15} for $\tau_3=1$. Accounting for the repeat-until-success (RUS) protocol, each synthesized rotation requires, on average, $5.4$ STAR rotations and $1.3~T$ gates, with $T^\dagger$ gates included in the $T$-gate count. In the RUS protocol, a failed STAR injection at angle $\phi$ is corrected by attempting a rotation at angle $2\phi$, with the correction angle doubled after each subsequent failure.

The second arbitrary rotation appearing in each local projector has angle
\begin{equation}
    \theta'=-\frac{\pi}{2}-\arctan\left(\frac{1}{\sqrt{2}}\right),
\end{equation}
and has the same STAR-rotation and $T$-gate costs, since
\begin{equation}
    R_Z(\theta')=R_Z\left(-\frac{\pi}{2}\right)R_Z(\theta),
\end{equation}
where $R_Z(-\pi/2)$ is a Clifford operation. Including the remaining gates in the circuits, the optimal detector therefore requires, on average, $13.7~T$ gates and $32.6$ STAR rotations, while the reduced detector requires $9.1$ $T$ gates and $21.7$ STAR rotations. These results are summarized in Table~\ref{tab:four-copy-comparison}.

\subsection{Code-level implementation and physical-qubit cost}
\label{sec:qec-implementation}

The resource counts above describe the logical implementation of the circuits. We finally consider the physical-qubit cost associated with encoding these logical operations. We use quantum Reed--Muller (QRM) codes and their punctured variants, focusing on the $[[7,1,3]]$ Steane code and the $[[8,3,2]]$ color code. The Steane code is a self-dual punctured QRM code that encodes one logical qubit per block, while the $[[8,3,2]]$ code, corresponding to $\mathrm{QRM}(0,1,3)$, encodes three logical qubits in a single block and can therefore accommodate one complete copy of our three-qubit state.

The Steane code provides the most direct implementation of the circuits considered here. Each of the twelve logical qubits is encoded in a separate $[[7,1,3]]$ block, giving a total of 84 physical data qubits. Since the Steane code supports transversal implementations of the logical Clifford generating set $\{H,S,\mathrm{CNOT}\}$, the Clifford operations appearing in the circuits of Figs.~\ref{fig:previous-four-copy} and~\ref{fig:new-circuits} can be implemented straightforwardly at the logical level.

To reduce the data-qubit overhead, we also consider the $[[8,3,2]]$ code. For the optimal detector shown in Fig.~\ref{fig:new-circuits}(b), we encode the four copies of the three-qubit state into four $[[8,3,2]]$ code blocks, with two copies initially encoded in the primal code and two in its dual. With this choice, all CNOT gates in the detector can be implemented transversally between code blocks. The first CNOT layer is implemented using permuted transversal CNOTs between dual and primal blocks. Transversal Hadamard gates then map the dual blocks to the primal code while simultaneously implementing the required logical Hadamards. Once all four blocks are in the primal code, the remaining CNOT gates are implemented using ordinary transversal CNOT layers. The corresponding code transformations and logical operations are detailed in Appendix~\ref{app:fault_tolerant_15circuit}.

To implement logical $T$ gates, we consider the preparation of high-fidelity $T$ magic states encoded in the $[[7,1,3]]$ code~\cite{itogawa2025efficient} and the $[[8,3,2]]$ code~\cite{chang2026smallest}. In the preparation protocols considered here, additional ancilla qubits are used for phase-kickback checks and stabilizer measurements, with a total of $2$--$4$ rounds of syndrome extraction. By contrast, the STAR factories considered here require only the encoded resource block and ancillas for stabilizer measurements, with a comparable number of syndrome-extraction rounds.

The total physical-qubit requirement depends on how many resource states must be supplied simultaneously. For the implementation of the Steane code, the reduced and optimal circuits require four and six concurrent non-Clifford operations, respectively, and we allocate one factory to each concurrent operation. For the $[[8,3,2]]$ implementation, we use the multi-output factory production rates of Ref.~\cite{chang2026smallest}, which reduce the number of simultaneous factories required for the optimal circuit. The corresponding physical-qubit requirements are summarized in Table~\ref{tab:qec_cost}. In the early fault-tolerant implementation, the additional STAR factories increase the instantaneous physical-qubit footprint, but the average number of resource-state injections is reduced by approximately a factor of four relative to direct Clifford+$T$ synthesis.

\begin{table}[t]
\caption{\textbf{Physical-qubit requirements for the fully and early fault-tolerant implementations.} We assume superdense syndrome extraction for the Steane code and flag-based syndrome extraction for the $[[8,3,2]]$ code. The phase-kickback check requires $6$ and $4$ additional physical qubits for factories based on the $[[7,1,3]]$ and $[[8,3,2]]$ codes, respectively. The $[[8,3,2]]$ totals use the factory production rates of Ref.~\cite{chang2026smallest}. N/A denotes cases not considered here.}
\label{tab:qec_cost}
\begin{ruledtabular}
\begin{tabular}{lcc}
&
$[[7,1,3]]$
&
$[[8,3,2]]$
\\
\midrule % \hline
$\#$ of data qubits
& $7\times12$
& $8\times4$
\\
$\#$ of readout qubits
& $6\times12$
& $8\times4$
\\
$\#$ of qubits for a $T$ factory
& $7+6+7$ \cite{itogawa2025efficient}
& $8+8+4$ \cite{chang2026smallest}
\\
$\#$ of qubits for a STAR factory
& $7+6$
& $8+8$
\\
Fig.~\ref{fig:new-circuits}(a) FTQC
& 236
& N/A
\\
Fig.~\ref{fig:new-circuits}(b) FTQC
& 276
& 124
\\
Fig.~\ref{fig:new-circuits}(a) Early FTQC
& 288
& N/A
\\
Fig.~\ref{fig:new-circuits}(b) Early FTQC
& 354
& 156
\end{tabular}
\end{ruledtabular}
\end{table}

\section{Discussion}

This work provides a simple example of how the abstract mathematical structure underlying a given quantum information task can be used to directly inform the design of a quantum experiment. In particular, the optimal four-copy detector is first identified from the quartic structure of the hyperdeterminant. Then, from a representation-theoretic perspective, the same measurement can be understood through the local $J=0$ sectors and the projection onto the trivial irreduciple representation of the relevant four-copy representation. This observation allowed us to decompose the corresponding local $J=0$ projectors into elementary gates and thus turn an abstract optimality result into a concrete circuit.

More broadly, our results suggest that finding a direct circuit for a nonlinear quantum quantity should not be viewed as the end of the problem. Once such a measurement is known, one can still ask how much of the available signal it captures and whether another measurement can extract the same information with fewer state preparations. Indeed, identifying the optimal measurement before compiling it into a circuit therefore provides a route for connecting the mathematical structure of quantum information with the practical resource requirements of quantum experiments.

Finally, the circuits introduced here may also be useful as test cases for early fault-tolerant devices. They are relatively small, but already combine transversal Clifford operations with non-Clifford rotations implemented through magic-state injection or STAR. Since the ideal click probabilities are known exactly, one can directly study how logical errors modify the statistical advantage of the different detectors. In particular, the $[[8,3,2]]$ implementation of the optimal detector requires only four encoded blocks, making it a relatively compact experiment for early fault-tolerant hardware.

\section{Acknowledgments}
We thank Maxwell West for useful discussions. This work was supported by Laboratory Directed Research and Development (LDRD) program of Los Alamos National Laboratory (LANL) under project number 20260043DR and by LANL's ASC Beyond Moore’s Law project. This material is based upon work supported by the U.S. Department of Energy, Office of Science, National Quantum Information Science Research Centers, Quantum Science Center (LC).

\medskip

\section{Artificial Intelligence Disclosure}
The idea for this project was created by humans. ChatGPT 5.6 Sol was used for literature review and for building the explicit circuits for the local $J=0$ sector projectors. All results were verified by the authors.

\bibliography{quantum}

\clearpage
\onecolumngrid
\appendix

\section{Proof of Proposition~\ref{prop:optimallity}}\label{app:proof-prop-1}

In this section we present a proof for Proposition~\ref{prop:optimallity}. 
First, let $\mathbb{P}(\mathcal{H})$ be the projective space associated with $\mathcal{H}$, whose points $[\phi]$ correspond to nonzero vectors $\ket{\phi}\in\mathcal{H}$ modulo multiplication by a nonzero complex number, also known as rays. 
Next, consider the set of pure three-qubit states with vanishing three-tangle
\begin{equation}
    \mathcal{W} =
    \left\{
    [\phi]\in\mathbb{P}(\mathcal{H})
    \;:\;
    \tau_3(\phi)=0
    \right\}\,,
\end{equation}
and define
\begin{equation}
    K_4(\mathcal{W}) = 
    {\rm span}
    \left\{
    \ket{\phi}^{\otimes 4}
    \;:\;
    [\phi]\in\mathcal{W}
    \right\}
    \subseteq
    \operatorname{Sym}^4(\mathcal{H}).
    \label{eq:K4W-proof}
\end{equation}
The orthogonal complement of $K_4(\mathcal{W})$, within the symmetric four-copy subspace, is in one-to-one correspondence with homogeneous quartic polynomials that vanish on $\mathcal{W}$.
Indeed, for every
$\ket{v}\in\operatorname{Sym}^4(\mathcal{H})$ we associate the quartic polynomial
\begin{equation}
    p_v(\psi)
    =
    \langle{v}\ket{\psi}^{\otimes 4}.
    \label{eq:riesz-polynomial-proof}
\end{equation}
Then $\ket{v}$ is orthogonal to $K_4(\mathcal{W})$ if and only if $p_v(\phi)=0$ for every $[\phi]\in\mathcal{W}$.

To determine its degree-four relations, let
$R=\mathbb C[a_{000},\ldots,a_{111}]$ and define the affine zero locus
\begin{equation}
    V(\Delta)
    =
    \left\{
    a\in\mathbb C^8
    \;:\;
    \Delta(a)=0
    \right\}.
\end{equation}
Since $\Delta$ is homogeneous, $V(\Delta)$ is the affine cone over
$\mathcal W$. We therefore write
\begin{equation}
    \hat{\mathcal W}=V(\Delta)\subset\mathbb C^8.
\end{equation}
Then, as $\Delta$ is irreducible in $R$, the principal ideal $(\Delta)$ is prime and radical.    
Application of Hilbert's Nullstellensatz~\cite{cox2015ideals} yields
    \begin{equation}
        I(\widehat{\cal W}) = \sqrt{(\Delta)} = (\Delta),
    \end{equation}
    where the last equality is a consequence of the radicality of $(\Delta)$.
    Due to the fact that $\Delta$ is homogeneous of degree four, taking the degree-four component yields 
    \begin{equation}
        I(\widehat{\mathcal W})_4 = {\mathbb C} \Delta.
    \end{equation}
    Thus, every homogeneous quartic polynomial vanishing on ${\cal W}$ is proportional to the hyperdeterminant $\Delta(\psi)$.

An immediate consequence of the previous argument establishes why four copies are necessary for nontrivial one-sided detection of the three-tangle (one can arrive at the same conclusion via representation theoretic arguments, we refer the reader to Appendix~\ref{app:rep-theory} for more details). Since, as previously established, $I(\widehat{\cal W}) = (\Delta)$ and $\Delta$ is homogeneous of degree four, its vanishing ideal has no nonzero homogeneous components of degree $n < 4$, 

\begin{equation}
    I(\widehat{\cal W})_n = \{0\}, \qquad n < 4.
\end{equation}

By the polynomial-symmetric-tensor correspondence, 
\begin{equation}
    K_n({\cal W}) = \operatorname{Sym}^n(\mathcal H),
    \qquad n<4.
\end{equation}
where $K_n(\mathcal W) ={\rm span}\{ \ket{\phi}^{\otimes n}:[\phi]\in\mathcal W\}$ generalizes \eqref{eq:K4W-proof}.
Thus, any positive $n$-copy effect with zero click probability on every state in ${\cal W}$ must annihilate $K_n({\cal W})$ and hence every $n$-identical copy pure state input.
This immediately rules out nontrivial universally one-sided detector for $n < 4$.

Next, we concern ourselves with proving that four is the minimum number of copies under the stated assumptions.
Using Eq.~\eqref{eq:delta-four-copy}, the polynomial $\Delta$ corresponds to the symmetric vector $\ket{w_\Delta}$. Hence the previous result implies
\begin{equation}
    K_4(\mathcal{W})^\perp
    \cap
    \operatorname{Sym}^4(\mathcal{H})
    =
    {\rm span}
    \left\{
    \ket{w_\Delta}
    \right\}.
    \label{eq:K4-complement-proof}
\end{equation}
Therefore, within the symmetric four-copy subspace ${\rm Sym}^{4}({\cal H})$, there is a unique direction, up to scalar multiplication, that is orthogonal to the fourth tensor power of every pure state with vanishing three-tangle.

We now consider an arbitrary effect $E$ satisfying the assumptions of the proposition.
Since $E\geq0$, for every $\ket{\phi}$ with $\tau_3(\phi)=0$ we have
\begin{equation}
    0
    =
    \bra{\phi}^{\otimes4}
    E
    \ket{\phi}^{\otimes4}
    =
    \left\|
    E^{1/2}\ket{\phi}^{\otimes4}
    \right\|^2.
\end{equation}
It follows that
\begin{equation}
    E^{1/2}\ket{\phi}^{\otimes4}=0,
\end{equation}
and therefore
\begin{equation}
    E\ket{\phi}^{\otimes4}=0.
\end{equation}
By linearity, $E$ annihilates the entire space $K_4(\mathcal{W})$.

Next, let
\begin{equation}
    \widetilde E = P_{\mathrm{sym}}^{(4)} E P_{\mathrm{sym}}^{(4)}
\end{equation}
be the restriction of $E$ to the symmetric four-copy subspace.
Since $K_4(\mathcal{W})\subseteq\operatorname{Sym}^4(\mathcal{H})$, the operator $\widetilde E$ also annihilates $K_4(\mathcal{W})$. 
Moreover, $\widetilde E\geq0$, so its support must lie in the orthogonal complement of $K_4(\mathcal{W})$.
From Eq.~\eqref{eq:K4-complement-proof}, this complement is one-dimensional. 
Thus
\begin{equation}
    \widetilde E =     \eta Q_4
\end{equation}
for some $\eta\geq0$, where
\begin{equation}
    Q_4
    =
    \frac{\ket{w_\Delta}\!\bra{w_\Delta}}
    {\langle w_\Delta|w_\Delta\rangle}
    =
    \frac{\ket{w_\Delta}\!\bra{w_\Delta}}{3}.
\end{equation}
Finally, $E\leq\mathbb{I}$ implies
\begin{equation}
    0
    \leq
    \widetilde E
    \leq
    P_{\mathrm{sym}}^{(4)}.
\end{equation}
Restricting this inequality to the one-dimensional support of $Q_4$ gives
$0\leq\eta\leq1$, and consequently,
\begin{equation}
    P_{\mathrm{sym}}^{(4)}
    E
    P_{\mathrm{sym}}^{(4)}
    =
    \eta Q_4
    \leq
    Q_4,
\end{equation}
which proves Eq.~\eqref{eq:optimal-effect-bound}.

For an arbitrary pure state $\ket{\psi}$, the vector
$\ket{\psi}^{\otimes4}$ lies in the symmetric subspace, and therefore
\begin{align}
    \bra{\psi}^{\otimes4}
    E
    \ket{\psi}^{\otimes4}
    &=
    \bra{\psi}^{\otimes4}
    P_{\mathrm{sym}}^{(4)}
    E
    P_{\mathrm{sym}}^{(4)}
    \ket{\psi}^{\otimes4}
    \nonumber\\
    &\leq
    \bra{\psi}^{\otimes4}
    Q_4
    \ket{\psi}^{\otimes4}
    \nonumber\\
    &=
    \frac{
    \left|
    \langle{w_\Delta}\ket{\psi}^{\otimes4}
    \right|^2
    }{3}
    \nonumber\\
    &=
    \frac{|\Delta(\psi)|^2}{3}
    =
    \frac{\tau_3(\psi)^2}{48}.
\end{align}
This proves Eq.~\eqref{eq:optimal-probability-bound}.

\section{Earlier four-copy constructions}
\label{app:prior}
In this section, we discuss the previous four-copy circuits for detecting the three-tangle. In particular, these are the protocols in Refs.~\cite{yu2007measurable}~and~\cite{yahyavi2022quantum}. Both proposals give a click probability of
\begin{equation}
    q_{\mathrm{prev}} = \frac{\tau_3^2}{256}.
\end{equation}
We include the derivations of these quantities.

In both protocols, four copies of a three-qubit quantum state are used, which we denote as $|\psi\rangle_1 \otimes |\psi\rangle_2 \otimes |\psi\rangle_3 \otimes |\psi\rangle_4$. Moreover, to keep track of which qubit belongs to whom, we use the standard A (Alice),  B (Bob) and C (Charlie) notation, to indicate first, second and third qubit, so that each copy carries the indexes
\begin{equation}
    |\psi\rangle_j = |\psi\rangle_{A_j B_j C_j} \in \mathcal{H}=\mathcal{H}_A\otimes \mathcal{H}_B\otimes \mathcal{H}_C
    \label{eq:subsystem_labels}
\end{equation}
for $j\in \{1,2,3,4\}$. In total, there are twelve physical qubits. 

Consider the projector onto the singlet state
\begin{equation}
    P = |s\rangle \langle s|, \quad \text{with} \quad |s\rangle = \frac{|01\rangle -|10\rangle}{\sqrt 2}. 
\end{equation}
The protocol in Ref~\cite{yu2007measurable} projects onto the singlet in qubit pairs 
\begin{equation}
    (A_1, A_2), (A_3, A_4), (B_1,B_2), (B_3,B_4), (C_1, C_3), (C_2, C_4).  
\end{equation}
The projector onto the singlet state between qubit $\alpha_i$ and $\alpha_j$ is given by 
\begin{equation}
    P_{ij}^\alpha = |s\rangle \langle s|_{\alpha_i \alpha_j},
\end{equation}
where $\alpha\in \{A, B,C\}$ and $i,j\in \{1,2,3,4\}$. Then, the total projector can be defined as
\begin{equation}\label{eq:projector_yu_song}
    \mathcal A:= P_{12}^A\,P_{34}^A\, P_{12}^B\,P_{34}^B\, P_{13}^C\, P_{24}^C = | S\rangle \langle S|,
\end{equation}
where $|S\rangle = |s\rangle_{A_1A_2}|s\rangle_{A_3A_4} |s\rangle_{B_1B_2}|s\rangle_{B_3B_4}|s\rangle_{C_1C_3}|s\rangle_{C_2C_4}$. 
The click event occurs when all six pair of projections and measurements return the singlet outcome. The probability of this event is given by 
\begin{equation}
    q_{\mathrm{prev}} = \langle \mathcal A \rangle = \langle \psi|^{\otimes 4}|S\rangle \langle S |\psi\rangle^{\otimes 4} = |\langle S|\psi\rangle^{\otimes 4}|^2.
\end{equation}
Considering that each singlet state acts on computational basis elements as
\begin{equation}
    \langle s|ij\rangle = \frac{\epsilon_{ij}}{\sqrt 2},
\end{equation}
where $\epsilon_{01}= 1$ and $\epsilon_{10} = -1$ and the other entries vanish. Writing $|\psi\rangle = \sum_{i,j,k =0}^1a_{ijk}|ijk\rangle$, the amplitude is 
\begin{equation}
    \langle S|\psi\rangle^{\otimes 4} = \frac 18 \sum \epsilon_{i_1 i_2} \epsilon_{i_3 i_4} \epsilon_{j_1 j_2} \epsilon_{j_3 j_4} \epsilon_{k_1 k_3} \epsilon_{k_2 k_4} \, a_{i_1 j_1 k_1} a_{i_2 j_2 k_2 } a_{i_3 j_3 k_3 } a_{i_4 j_4 k_4 }
\end{equation}
where the factor $1/8$ comes from the six normalized singlets. Directly expanding the epsilon contractions and collecting identical products of amplitudes gives
\begin{equation}
    \langle S|\psi\rangle^{\otimes 4}  = -\frac 14 \Delta(\psi),
\end{equation}
where $\Delta(\psi)$ is Cayley’s hyperdeterminant. Identifying $\tau_3 (\psi) = 4 |\Delta(\psi)|$ and computing the click probability leads to 
\begin{equation}
   q_{\mathrm{prev}} = \langle \mathcal A\rangle = \frac{\tau_3^2}{256}. 
\end{equation}

Each singlet projection appearing in $\mathcal A$ can be implemented destructively using a Bell-basis measurement. For a pair of qubits, apply a CNOT with the first qubit as the control, followed by a Hadamard on the first qubit. Both qubits are then measured in the computational basis. Denoting this two qubit circuit by 
\begin{equation}
    U_{\operatorname{Bell}} = (H\otimes \mathbb I) CNOT,
\end{equation}
its action on the singlet state is $U_{\operatorname{Bell}} |s\rangle = |11\rangle$. It follows that 
\begin{equation}
    |s\rangle \langle s | = U_{\operatorname{Bell}} ^\dagger |11\rangle \langle 11| U_{\operatorname{Bell}}.
\end{equation}
Therefore, the probability of obtaining the computational-basis outcome $|11\rangle$ after this circuit is exactly the probability that the input state lies on the singlet subspace. 

This construction is applied to the six disjoint pairs in Eq.~\eqref{eq:projector_yu_song}, as shown in Fig.~\ref{fig:previous-four-copy}(a). Let $\mathcal U_{\operatorname{Bell}}$ denote the product of each $U_{\operatorname{Bell}}$ applied to the corresponding pairs. Then 
\begin{equation}
    \mathcal A = \mathcal U_{\operatorname{Bell}}^\dagger(|11\rangle\langle 11|)^{\otimes 6}\mathcal U_{\operatorname{Bell}}.
\end{equation}
The click event occurs when all twelve physical qubits are measured in the state $|1\rangle$. The click probability is therefore 
\begin{equation}
   q_{\mathrm{prev}} = |\langle 11|^{\otimes 6}\mathcal U_{\operatorname{Bell}}|\psi\rangle^{\otimes 4}|^2 = \frac{\tau_3(\psi)^2}{256}.
\end{equation}

Ref.~\cite{yahyavi2022quantum} presents an alternative circuit that yields the same click probability as Ref.~\cite{yu2007measurable}. To maintain the subsystem-and-copy notation introduced in Eq.~\eqref{eq:subsystem_labels}, we identify the numerical qubit labels used in
Ref.~\cite{yahyavi2022quantum} according to
\begin{equation}
(1,2,3) =(A_1,B_1,C_1), \quad
(4,5,6) =(A_2,B_2,C_2), \quad
(7,8,9) =(A_3,B_3,C_3), \quad
(10,11,12) =(A_4,B_4,C_4).
\end{equation} 

The circuit first applies single-qubit Pauli $Y$ gates on qubits $A_2,B_2,C_3,A_4,B_4$ and $C_4$, followed by the five controlled-$X$ gates 
\begin{equation}
\mathrm{CNOT}_{B_1\to B_2}\mathrm{CNOT}_{C_1\to C_3}\mathrm{CNOT}_{C_2\to C_4}\mathrm{CNOT}_{A_3 \to A_4}\mathrm{CNOT}_{B_3\to B_4},
\label{eq:yahyavi_cnots}
\end{equation}
where we denote $\mathrm{CNOT}_{i\to j}$ the controlled-$X$ gate conditioning on the $i$th qubit, with qubit $j$ as the target. Since the five CNOT gates act on pairwise disjoint qubit pairs, they can be applied in parallel and have entangling depth one at the logical circuit level. 

Finally, the circuit is followed by the single-qubit gate $R = R_Y(-\pi/2)$ on qubits $B_1,C_1,C_2,A_3$ and $B_3$. The gate $R$ transforms single-qubit state as: 
\begin{equation}
R\ket0=\frac{\ket0-\ket1}{\sqrt2},
\qquad
R\ket1=\frac{\ket0+\ket1}{\sqrt2}.
\end{equation}

Let $U_{\rm YJHK}$ denote the unitary implemented by the complete circuit, which is summarized in Fig.~\ref{fig:previous-four-copy}(b).
At the end of the transformation applied to the four copies of the initial state yields the final state 
\begin{equation}
    \ket{\eta}  = U_{\rm YJHK} \ket{\psi}^{\otimes 4} =  -\frac{\Delta(\psi)}{4\sqrt{2}}(\ket{s_0} + \ket{s_1}) + \ket{\eta_{\perp}}\,,
\end{equation}
where $\ket{s_0} = (\ket{000})^{\otimes 4}$ and $\ket{s_1} = (\ket{100})^{\otimes 2 } (\ket{000})^{\otimes 2}$ are the computational basis states whose amplitudes are proportional to the hyperdeterminant. The state  $\ket{\eta_{\perp}}$    contains the remaining
computational-basis components and satisfies
$\langle s_0|\eta_{\perp}\rangle =\langle s_1|\eta_{\perp}\rangle=0$.  

Using $\tau_3(\psi)=4|\Delta(\psi)|$, each of the two relevant outcomes occurs with probability  $\mathbb{P}(s_0)  = \mathbb{P}(s_1) = \tau_3^2/512$~\cite{yahyavi2022quantum}. Coarse graining of the two outcomes into a single detection event gives
\begin{equation}
q_{\rm click}=  \mathbb{P}(s_0) + \mathbb{P}(s_1)  = \frac{\tau_3^2}{256}.
\label{eq:yahyavi_prob}
\end{equation}

\section{Proof of proposition~\ref{prop:circuits}}\label{app:proof-prop-2}

\subsection{The 15-CNOT circuit}

We derive the circuit for the local $J=0$ measurement shown in Fig.~\ref{fig:new-circuits}(a). For convenience, we group the four copies by qubit, and run the projector per qubit $X=A,B,C$ across  $X_1 X_2 X_3 X_4$.
For each pair of copies, define
\begin{equation}
\ket{s}=\frac{\ket{01}-\ket{10}}{\sqrt2},\quad
\ket{t_+}=\ket{00},\quad
\ket{t_0}=\frac{\ket{01}+\ket{10}}{\sqrt2},\quad
\ket{t_-}=\ket{11}.
\end{equation}
This gives rise to the orthonormal basis for the four-spin $J=0$ sector in  Eq.~\eqref{eq:logical}, namely
\begin{align}
\ket{0_L}&=\ket{s}_{12}\ket{s}_{34},\nonumber\\
\ket{1_L}&=\frac{1}{\sqrt3}
\left(
\ket{t_+}_{12}\ket{t_-}_{34}
-\ket{t_0}_{12}\ket{t_0}_{34}
+\ket{t_-}_{12}\ket{t_+}_{34}
\right),
\label{eq:logical}
\end{align}
Recall that we denote the projection onto the $J=0$ with $P_0=|0_L\rangle\langle0_L| + |1_L\rangle\langle 1_L|$, and that the goal is to find a circuit that corresponds to measuring $Q_4 = P_0^A\otimes P_0^B\otimes P_0^C$. This means we can execute three measurements of $P_0^X$ for $X=A,B,C$ and combine the result with a classical logical AND operation to obtain $Q_4$.

To that end, we choose to find a circuit that maps the $J=0$ sector spanned by $\ket{0_L}, \ket{1_L}$ to ${\rm span}\{\ket{1100},\ket{0000}\}$ to facilitate measurement in the computational basis. This way, we may see the $J=0$ projection as a stabilizer code generated by $\langle Z_1Z_2, Z_3, Z_4\rangle$, that is, we accept a measurement outcome if even parity across the first two qubits and zero on the last two qubits. 

Let $\mathrm{CNOT}_{i\to j}$ denote conditioning on the $i$th qubit, with qubit $j$ as the target.
The first stage of the circuit applies the inverse Bell transform to pairs $(1,2)$ and $(3,4)$, moving into the computational basis. In our convention this means a CNOT conditioned on the first acting on the second qubit of each pair followed by a Hadamard gate on the first qubit.  Subsequently, we apply $\mathrm{CNOT}_{1\to3}$ and $\mathrm{CNOT}_{2\to4}$. We summarize these operations by 
\begin{equation}
    U_4 = 
    \big(\mathrm{CNOT}_{1\to 3} \mathrm{CNOT}_{2\to 4}\big)
    H_1 H_3 
    \big(\mathrm{CNOT}_{1\to 2} \mathrm{CNOT}_{3\to 4}\big)
\end{equation}
Direct substitution gives
\begin{align}
U_4\ket{0_L}&=\ket{1100},\nonumber\\
U_4\ket{1_L}&=\frac{1}{\sqrt3}
\left(\ket{0000}-\ket{0100}-\ket{1000}\right).
\label{eq:after4}
\end{align}
The last two physical qubits are therefore fixed to $\ket{00}_{34}$ throughout the image of the $J=0$ sector. In the next step, we rotate the two-dimensional subspace of qubits $1,2$ into the even-parity subspace ${\rm span}\{\ket{00}_{12},\ket{11}_{12}\}$.
Set
\begin{equation}
\theta=\arctan\left(\frac{1}{\sqrt2}\right)\,,
\end{equation}
and define the two-qubit unitary
\begin{equation}
V= \big(R_y(-\theta)\otimes R_y(-\pi/2-\theta)\big)
\; \mathrm{CNOT}_{1\to2}\; \big(R_y(-3\pi/4)\otimes R_y(-\pi/4)\big).
\label{eq:V}
\end{equation}
Acting with $V$ on the first two qubits in Eq.~\eqref{eq:after4} gives
\begin{align}
    (V\otimes I_{34})U_4\ket{0_L}
        &=\cos(\frac{\pi}{12})\ket{0000}
        -\sin{\frac{\pi}{12}}\ket{1100},\nonumber\\
        (V\otimes I_{34})U_4\ket{1_L}
        &=-\sin(\frac{\pi}{12})\ket{0000}
        -\cos(\frac{\pi}{12})\ket{1100}.
\label{eq:decoder_images}
\end{align}
Hence, the output states of the circuit are in ${\rm span}\{ \ket{1100}, \ket{0000} \}$ and thereby measurement in the computational basis is equivalent to measuring $P_0$.

\subsection{The 11-CNOT circuit}
\label{app:reduced}

In this section we outline how to achieve a reduction from 15~CNOTs required for the circuit in Fig.~\ref{fig:new-circuits}(a) to 11~CNOTs.
The normalized vector $\ket{w_\Delta}/\sqrt{3}$ arising from the definition of the hyperdeterminant in Eq.~\eqref{eq:delta-four-copy} projected onto the logical singlet/triplet basis from Eq.~\eqref{eq:logical} gives 
\begin{equation}
\ket{\Omega}
=\frac12\left(
-\ket{0_L0_L1_L}
-\ket{0_L1_L0_L}
-\ket{1_L0_L0_L}
+\ket{1_L1_L1_L}
\right)\,.
\label{eq:omega_full}
\end{equation}
Tracing out any two logical qubits gives
\begin{equation}
\rho_L^{A}=\rho_L^{B}=\rho_L^{C}=\frac{I_L}{2}.
\label{eq:maxmixed}
\end{equation}
Within a local $J=0$ subspace, a singlet projection on physical qubits $1,2$ is  simply
\begin{equation}
P_{12}^-\big|_{J=0}=\proj{0_L},
\label{eq:singlet_logical}
\end{equation}
where an implementation to map the first two qubits onto the even-parity subspace ${\rm span}\{\ket{11}, \ket{00}\}$ is simply $H_1 \mathrm{CNOT}_{1\to 2}$.
Consider therefore
\begin{equation}
E_{11}=P_0^A\otimes P_0^B\otimes P_{12}^{-,C}.
\label{eq:E11}
\end{equation}
Then, at the cost of 5 CNOT gates each, we accept on $A, B$ if  the outcome is 0000 or  1100. Additionally, at the cost of one CNOT, we accept C if the first two qubits are 11. We know from the section above that on the $J=0$ sector, the last two qubits are always 00. Therefore, intuitively, we skip the $\ket{00}_{12,C}$ outcomes that would be counted under the $P_0^A \otimes P_0^B\otimes P_0^C$ construction and only accept the $\ket{11}_{12,C}$ outcomes, which would amount to doubling the number of clicks necessary.  More formally, this is justified by Eq.~\eqref{eq:maxmixed}, as measuring on qubit $C$ with $P_{12}^{-,C}\otimes I_{34}$ amounts to tracing out the last two qubits.
Equation~\eqref{eq:singlet_logical} then keeps half of its norm. Therefore
\begin{equation}
P_{\rm sym}E_{11}P_{\rm sym}=\frac12 Q_4,
\end{equation}
which proves Eq.~\eqref{eq:q11}.

\section{Representation-theoretic origin of the $J=0$ measurement}
\label{app:rep-theory}

As mentioned in the main text, we can directly understand the three-tangle and how to compute it by projecting onto a  local $J=0$ sector directly from representation theory. In particular, we note that the discussion presented here ties to the results in~\cite{leifer2004measuring,yu2007measurable} which construct the three-tangle in terms of contractions with the antisymmetric tensor $\epsilon$. Since $\epsilon$ is the unique two-qubit singlet tensor, these contractions naturally live in the total-spin-zero sectors of the four local copies.

To begin, it is also useful to note why four copies are necessary in the first place. In particular, one can readily show that there are no nonconstant homogeneous $SL(2,\mathbb C)^{\otimes 3}$ invariants of degree smaller than four. For one and three copies this follows immediately because an odd number of spin-$1/2$ representations contains no local $J=0$ sector. At degree two, each party has a unique singlet, but the product of the three local singlets is antisymmetric under exchange of the two copies and therefore has zero overlap with the globally symmetric state $\ket{\psi}^{\otimes 2}$. The first nonvanishing invariant can thus occur at degree four, where the trivial representation appears in the relevant product of local $J=0$ sectors.

More precisely, consider the Hilbert space associated to all of Alice's qubits (i.e., the space of the four copies of the first qubit). From the Schur--Weyl duality we know that~\cite{bacon2006efficient}
\begin{equation}
    \mathcal{H}_A^{\otimes 4}=\left(\mathbb C^2\right)^{\otimes 4}
    \simeq
    V_{2}\otimes[4]
    \oplus
    V_{1}\otimes[31]
    \oplus
    V_{0}\otimes[22],
    \label{eq:schur-four-qubits}
\end{equation}
where $V_J$ denotes the spin-$J$ irreducible representation of $SU(2)$ and $[\lambda]$ denotes an irreducible representation of the copy-permutation group $S_4$. The local $J=0$ sector is therefore two dimensional and carries the $[22]$ representation of $S_4$. Clearly, a similar decomposition will hold for the Hilbert space of Bob and Charlie's qubits. To differentiate between the parties, we will denote their corresponding projectors onto the $J=0$ subspace as $P_0^X$, with  $X=A,B,C$.

As per Eq.~\eqref{eq:transform}, the vector $\ket{w_\Delta}$ associated with the hyperdeterminant is invariant under the collective action of $SL(2,\mathbb C)$ on the four copies of each local qubit. In particular, it is invariant under the corresponding $SU(2)$ action and hence satisfies
\begin{equation}
    \left(
    P_0^A\otimes P_0^B\otimes P_0^C
    \right)
    \ket{w_\Delta}
    =
    \ket{w_\Delta}.
    \label{eq:wdelta-j0}
\end{equation}

It remains to determine how much of the product $J=0$ sector is accessible to four identical copies. Since $\ket{\psi}^{\otimes4}$ is invariant under simultaneous permutations of the copies, it lies in the trivial $[4]$ representation of the diagonal action of $S_4$. Moreover, one can verify that the trivial representation of $S_4$ occurs with multiplicity one in
\begin{equation}
    [22]\otimes[22]\otimes[22].
\end{equation}
Consequently, and given that $P_0^A\otimes P_0^B\otimes P_0^C$ commutes with the $S_4$ action, it follows that the product of the three local $J=0$ projectors has rank one when restricted to the globally symmetric four-copy subspace. Since $\ket{w_\Delta}$ lies in this subspace and in the product $J=0$ sector, this unique direction must be the hyperdeterminant direction. Therefore
\begin{equation}
    P_{\rm sym}^{(4)}
    \left(
    P_0^A\otimes P_0^B\otimes P_0^C
    \right)
    P_{\rm sym}^{(4)}
    =
    Q_4
    =
    \frac{\ket{w_\Delta}\!\bra{w_\Delta}}{3}.
    \label{eq:j0-equals-Q4}
\end{equation}

The previous result is exactly the basis for our circuits. Indeed, given four copies, of the state of interest, we can simultaneously project all local parties onto $J=0$. Its click probability is therefore
\begin{equation}
    \bra{\psi}^{\otimes4}
    P_0^A\otimes P_0^B\otimes P_0^C
    \ket{\psi}^{\otimes4}
    =
    \frac{\tau_3(\psi)^2}{48}.
\end{equation}
\section{Details for fault-tolerant implementation}
\label{app:fault-tolerant}

In this section we prove Proposition~\ref{prop:clifford-t-cost} and Corollary~\ref{cor:ft-precision}. We first derive the Clifford+$T$ cost of the circuits in Fig.~\ref{fig:new-circuits}, and then bound the effect of the approximate rotation synthesis on their click probabilities.

\subsection{Proof of Proposition~\ref{prop:clifford-t-cost}}

We begin with the previous four-copy constructions. The circuit of Ref.~\cite{yu2007measurable} consists of six destructive Bell-basis measurements. As discussed in Appendix~\ref{app:prior}, each of these measurements can be implemented using one CNOT and one Hadamard gate. The complete circuit is therefore Clifford and requires no $T$ gates. Similarly, the circuit of Ref.~\cite{yahyavi2022quantum} contains Pauli $Y$ gates, CNOT gates, and the rotation
\begin{equation}
    R_y(-\pi/2),
\end{equation}
all of which are Clifford operations. Hence, both previous constructions have zero $T$-count.

We next consider the local $J=0$ projector used in our circuits. Its nontrivial two-qubit part is given by Eq.~\eqref{eq:V}, that is,
\begin{equation}
    V
    =
    \left(
    R_y(-\theta)
    \otimes
    R_y(-\pi/2-\theta)
    \right)
    \mathrm{CNOT}
    \left(
    R_y(-3\pi/4)
    \otimes
    R_y(-\pi/4)
    \right),
    \label{eq:V-ft-appendix}
\end{equation}
where
\begin{equation}
    \theta
    =
    \arctan\left(\frac{1}{\sqrt{2}}\right).
\end{equation}
Since rotations about the same axis commute, we can write
\begin{equation}
    R_y(-\pi/2-\theta)
    =
    R_y(-\pi/2)R_y(-\theta).
    \label{eq:split-theta}
\end{equation}
Importantly, the first factor in Eq.~\eqref{eq:split-theta} is Clifford. Moreover,
$R_y(-\pi/4)$ and $R_y(-3\pi/4)$ are Clifford-equivalent to a single
$T$ or $T^\dagger$ gate~\cite{giles2013exact}. Thus, each local $J=0$ projector contains two exact
$T$ gates and two occurrences of the arbitrary rotation $R_y(-\theta)$.

To synthesize the latter, we use the fact that a $y$ rotation is related to a $z$ rotation by Clifford conjugation as $R_y(\alpha) =S H R_z(\alpha) H S^\dagger$. Then, since conjugation by Clifford gates preserves the operator norm, approximating
$R_z(-\theta)$ to precision $\epsilon$ gives an approximation of
$R_y(-\theta)$ with the same precision. Let $\#_{\rm GS}(\epsilon)$ denote
the $T$-count produced by \textsc{gridsynth} for this approximation.

The 11-CNOT circuit contains two local $J=0$ projectors and one singlet measurement. The singlet measurement is Clifford, and therefore
\begin{equation}
    \#_{T}^{(11)}
    =
    4+4\#_{\rm GS}(\epsilon).
    \label{eq:Tcount11-exact}
\end{equation}
The optimal 15-CNOT circuit contains three local $J=0$ analyzers, giving
\begin{equation}
    \#_{T}^{(15)}
    =
    6+6\#_{\rm GS}(\epsilon).
    \label{eq:Tcount15-exact}
\end{equation}
For typical instances, \textsc{gridsynth} achieves asymptotic scaling~\cite{ross2014optimal}
\begin{equation}
    \#_{\rm GS}(\epsilon)
    =
    3\log_2(1/\epsilon)
    +
    \mathcal{O}\!\left(
    \log\log(1/\epsilon)
    \right).
    \label{eq:gridsynth-scaling-app}
\end{equation}
Substituting Eq.~\eqref{eq:gridsynth-scaling-app} into
Eqs.~\eqref{eq:Tcount11-exact} and~\eqref{eq:Tcount15-exact} gives
\begin{align}
    \#_{T}^{(11)}
    =
    12\log_2(1/\epsilon)
    +
    \mathcal{O}\!\left(
    \log\log(1/\epsilon)
    \right)
    \qquad \text{and}\qquad
    \#_{T}^{(15)}
    =
    18\log_2(1/\epsilon)
    +
    \mathcal{O}\!\left(
    \log\log(1/\epsilon)
    \right),
\end{align}
which proves Proposition~\ref{prop:clifford-t-cost}.

\subsection{Proof of Corollary~\ref{cor:ft-precision}}

We now determine how synthesis error modifies the click probability. Let $U$ denote either of our exact circuits and let $\widetilde U$ denote the circuit obtained after replacing each arbitrary rotation by an approximation in operator norm satisfying
\begin{equation}
    \left\|
    R_y(\pm\theta)
    -
    \widetilde R_y(\pm\theta)
    \right\|
    \leq
    \epsilon.
    \label{eq:single-rotation-error}
\end{equation}
If the circuit contains $k$ such rotations, a telescoping expansion of the product of gates leads to
\begin{equation}
    \left\|
    U-\widetilde U
    \right\|
    \leq
    k\epsilon.
    \label{eq:unitary-telescoping}
\end{equation}
Let $\Pi$ denote the computational-basis acceptance projector associated with the corresponding detector. The exact ($E$) and synthesized ($\widetilde E$) operators are
\begin{equation}
    E
    =
    U^\dagger\Pi U,
    \qquad
    \widetilde E
    =
    \widetilde U^\dagger\Pi\widetilde U.
\end{equation}
Using $\|\Pi\|=1$ and the triangle inequality,
\begin{align}
    \left\|
    E-\widetilde E
    \right\|
    &\leq
    \left\|
    (U^\dagger-\widetilde U^\dagger)
    \Pi U
    \right\|
    +
    \left\|
    \widetilde U^\dagger
    \Pi
    (U-\widetilde U)
    \right\|
    \leq
    2
    \left\|
    U-\widetilde U
    \right\|
    \leq
    2k\epsilon.
    \label{eq:effect-error-bound}
\end{align}
Therefore, for any input state $\ket{\psi}$,
\begin{equation}
    \left|
    q(\psi)-\widetilde q(\psi)
    \right|
    \leq
    2k\epsilon.
    \label{eq:prob-error-bound}
\end{equation}
For the 11-CNOT circuit, there are $k=4$ arbitrary rotations. Its ideal click probability exceeds that of Ref.~\cite{yu2007measurable} by
\begin{align}
    q_{11}(\psi)-q_{\rm prev}(\psi)
    =
    \frac{\tau_3(\psi)^2}{96}
    -
    \frac{\tau_3(\psi)^2}{256}
    =
    \frac{5}{768}\tau_3(\psi)^2.
    \label{eq:gap11}
\end{align}
Equation~\eqref{eq:prob-error-bound} guarantees that the synthesized circuit still has click probability at least as large as the Clifford construction whenever
\begin{equation}
    8\epsilon
    \leq
    \frac{5}{768}\tau_3(\psi)^2.
\end{equation}
Hence it is sufficient that
\begin{equation}
    \epsilon
    \leq
    \frac{5}{6144}\tau_3(\psi)^2.
    \label{eq:epsilon11-app}
\end{equation}
For the optimal circuit, there are $k=6$ arbitrary rotations and
\begin{align}
    q_{\rm opt}(\psi)-q_{\rm prev}(\psi)
    =
    \frac{\tau_3(\psi)^2}{48}
    -
    \frac{\tau_3(\psi)^2}{256}
    =
    \frac{13}{768}\tau_3(\psi)^2.
    \label{eq:gap15}
\end{align}
In this case, Eq.~\eqref{eq:prob-error-bound} yields
\begin{equation}
    12\epsilon
    \leq
    \frac{13}{768}\tau_3(\psi)^2,
\end{equation}
and therefore
\begin{equation}
    \epsilon
    \leq
    \frac{13}{9216}\tau_3(\psi)^2.
    \label{eq:epsilon15-app}
\end{equation}
Equations~\eqref{eq:epsilon11-app} and~\eqref{eq:epsilon15-app} prove Corollary~\ref{cor:ft-precision}.

\subsection{Clifford and $T$-gate counts}
Here, we estimate the total number of logical Clifford operations to accompany the $T$-gate counts in Proposition~\ref{prop:clifford-t-cost}. We count each logical CNOT as one Clifford operation and each maximal single-qubit Clifford block as one Clifford operation. 

Let $\#_{\rm C,GS}(\epsilon)$ denote the number of single-qubit Clifford blocks generated when synthesizing one arbitrary rotation. The standard Clifford+$T$ normal form used by \textsc{gridsynth} contains approximately one Clifford block per $T$ gate~\cite{ross2014optimal,giles2013exact}, so that
\begin{equation}
    \#_{\rm C,GS}(\epsilon)
    =
    \#_{\rm GS}(\epsilon)
    +
    \mathcal{O}(1).
\end{equation}
The Clifford conjugations required to transform $R_z(\theta)$ into $R_y(\theta)$ can be absorbed into the endpoint Clifford blocks and therefore do not modify this scaling.

Each local $J=0$ projector contains five CNOT gates, two Hadamard gates, one $R_y(-\pi/2)$ Clifford rotation, two exact $T$ gates, and four additional Clifford blocks associated with the two $\pi/4$-type rotations. The resulting estimated resource counts are therefore
\begin{align}
    \#_{T}^{(11)}
    &=
    4+4\#_{\rm GS}(\epsilon),
    &
    \#_{\rm C}^{(11)}
    &=
    26+4\#_{\rm C,GS}(\epsilon),
    \\
    \#_{T}^{(15)}
    &=
    6+6\#_{\rm GS}(\epsilon),
    &
    \#_{\rm C}^{(15)}
    &=
    36+6\#_{\rm C,GS}(\epsilon).
\end{align}
For $\tau_3=1$, using the synthesis precisions required to retain the statistical advantage over Ref.~\cite{yu2007measurable}, we obtain the estimates
\begin{equation}
    \big(\#_{\rm C}^{(11)},\#_{T}^{(11)}\big)
    \sim
    (150,127),
    \qquad
    \big(\#_{\rm C}^{(15)},\#_{T}^{(15)}\big)
    \sim
    (210,177).
\end{equation}
By comparison, the circuits of Refs.~\cite{yu2007measurable,yahyavi2022quantum} are entirely Clifford. With the same counting convention, their resource counts are
\begin{equation}
\begin{aligned}
    \big(\#_{\mathrm C},\#_{T}\big)_{\text{Ref.~\cite{yu2007measurable}}}
    &= (12,0),\\
    \big(\#_{\mathrm C},\#_{T}\big)_{\text{Ref.~\cite{yahyavi2022quantum}}}
    &= (16,0).
\end{aligned}
\end{equation}

\subsection{No-go for an exact implementation of the optimal detector}
For completeness, we also justify why the non-Clifford synthesis cost of the optimal detector cannot be removed by finding a different exact Clifford+$T$ decomposition. Consider the GHZ state, for which $\tau_3=1$. Proposition~\ref{prop:optimallity} requires an optimal detector to click with probability
\begin{equation}
    q_{\rm opt}(\mathrm{GHZ})
    =
    \frac{1}{48}.
    \label{eq:GHZ-probability-app}
\end{equation}
The state $\ket{\mathrm{GHZ}}^{\otimes4}$ is a stabilizer state. For a finite Clifford+$T$ circuit, all amplitudes generated from a stabilizer input belong to the ring~\cite{giles2013exact} $
    \mathbb{Z}
    \left[
    \frac{1}{\sqrt{2}},i
    \right]$. Consequently, probabilities of computational-basis events belong to $
    \mathbb{Z}
    \left[
    \frac{1}{2},\sqrt{2}
    \right]$. If such a probability is rational, the coefficient multiplying $\sqrt{2}$ must vanish, and its denominator in lowest terms is therefore a power of two. However, for our optimal detector we have
\begin{equation}
    \frac{1}{48}
    \notin
    \mathbb{Z}\left[\frac{1}{2}\right].
\end{equation}
Thus, no finite exact Clifford+$T$ circuit with stabilizer ancillas, computational-basis readouts and no postelection can realize the optimal four-copy detector.

\section{Details for early fault-tolerant implementation}\label{app:early-fault-tolerant}

\subsection{Description on Quantum Reed Muller code \label{app:qrm_codes}}
In this appendix, we briefly review quantum Reed--Muller (QRM) codes and closely related punctured constructions. This family includes the $[[8,3,2]]$ and $[[16,6,4]]$ codes used in our fault-tolerant circuit implementations, while a punctured construction yields the $[[7,1,3]]$ Steane code.

We begin with the classical binary Reed--Muller code $\operatorname{RM}(r,m)$. Let
\begin{equation}
    \operatorname{Eval}(f) \coloneqq \left(f(\mathbf{a})\right)_{\mathbf{a} \in \mathbb{F}_2^m} 
\end{equation}
denote the $2^m$-dimensional vector obtained by evaluating a polynomial $f$ at all points of $\mathbb{F}_2^m$. The Reed--Muller code of order $r$ in $m$ variables is then defined as 
\begin{equation}
    \operatorname{RM}(r,m) := \left\{ \operatorname{Eval}(f): f\in \mathbb{F}_2[x_1, x_2, \dots, x_m] \;  \middle | \;\mathrm{deg}(f) \le r \right\}, 
\end{equation}
where $\mathbb{F}_2[x_1, x_2, \dots, x_m]$ denotes the polynomials with variables $x_1,\dots, x_m$ defined on the field $\mathbb{F}_2$. For $0\leq r\leq m$, this is a binary linear code with parameters  $[n = 2^m, k = \sum_{ i=0}^r\binom{m}{i}, d = 2^{m-r}]$.  
For $0\leq r<m$, its dual code with respect to the standard binary inner product is itself a Reed--Muller code, given as $ \operatorname{RM}(r,m)^\perp = \operatorname{RM}(m-r-1,m)$. 

Quantum Reed--Muller code  $\operatorname{QRM}(r_x, r_z, m)$ can be constructed as a Calderbank–Shor–Steane (CSS) code from pairs of classical RM codes, $\operatorname{RM}(r_x, m)$ and $\operatorname{RM}(r_z, m)$, which separately give $X$- and $Z$- stabilizer spaces.  The CSS commutation condition is satisfied provided that $r_x+r_z\leq m-1$, since $\operatorname{RM}(r_z, m)$ is a subset of the dual code $\operatorname{RM}(r_z,m) \subseteq \operatorname{RM}(r_x, m)^\perp = \operatorname{RM}(m - r_x - 1, m)$, or equivalently $\operatorname{RM}(r_x,m) \subseteq  \operatorname{RM}(r_z, m)^\perp$. 

The resulting code uses $n = 2^m$ physical qubits to encode $k = \sum_{i= r_x + 1}^{m - r_z - 1} \binom{m}{i}$ logical qubits, with distance $ d=\min(2^{r_x + 1}, 2^{r_z + 1})$. 
Representatives of the logical Pauli $X$- and $Z$-operators may be chosen from the quotient spaces,  $\operatorname{RM}(m-r_{z} - 1, m) / \operatorname{RM}(r_x, m)$ and  $\operatorname{RM}(m-r_{x} - 1, m) / \operatorname{RM}(r_z, m)$, respectively. 

The dual code is obtained by exchanging the $X$- and $Z$-stabilizer spaces. Accordingly, a transversal physical Hadamard on all physical qubits maps
\begin{equation}
\operatorname{QRM}(r_x,r_z,m)
\xrightarrow{H^{\otimes n}}
\operatorname{QRM}(r_z,r_x,m). 
\label{eq:qrm_hadamards}
\end{equation}
The precise induced action on the encoded logical qubits depends on the choice and labeling of logical operators and may include a permutation of the logical qubit labels~\cite{haah2017magic}.
Thus, $\operatorname{QRM}(r_z,r_x,m)$ is the dual of $\operatorname{QRM}(r_x,r_z,m)$, and the code is self-dual when $r_x=r_z$~\cite{nelson2025small}.

Several well-known quantum error correcting codes belong to this family or arise from closely related punctured constructions. For example, the $[[8,3,2]]$ code corresponds to $\operatorname{QRM}(0,1,3)$, while the tesseract code $[[16,6,4]]$ is $\operatorname{QRM}(1,1,4)$. Here, the dual code of $[[8,3,2]]$ codes remains $[[8,3,2]]$ but with the role of $X$ and $Z$ stabilizers exchanged, in other words, the dual code $\operatorname{QRM}(0, 1, 3)$ corresponds to $\operatorname{QRM}(1, 0, 3)$. 
The $[[7,1,3]]$ Steane code is obtained from punctured and shortened versions of the first-order Reed--Muller code with $m=3$ and therefore is regarded as a punctured QRM code. 

\begin{figure}
    \subfloat[]{\includegraphics[width=0.25\linewidth]{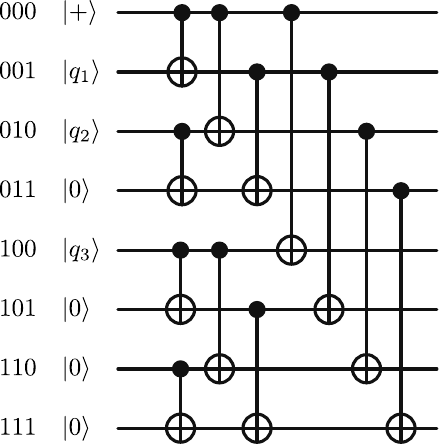}}
    \hspace{3cm}
    \subfloat[]{\includegraphics[width=0.25\linewidth]{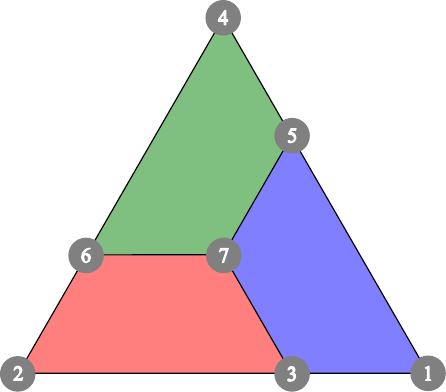}}
    
    \caption{\textbf{Summary of quantum reed--muller code and the methods used in this paper for fault-tolerant architecture.} (a) Hypercube circuit for encoding an arbitrary state into a QRM code. The example shown encodes the three-qubit state $\ket{q_1,q_2,q_3}$ into the $[[8,3,2]]$ code. (b) The $[[7,1,3]]$ Steane code. Gray circles represent physical qubits, while the green, red, and blue faces indicate the supports of the corresponding $X$- and $Z$-type stabilizer generators. }
    \label{fig:qrm_codes}
\end{figure}

An arbitrary $k$-qubit state $\ket{\psi}$ can be encoded into the logical subspace of QRM code using the hypercube encoding circuit shown in Fig.~\ref{fig:qrm_codes}(a)~\cite{gong2024computation}. In this circuit, each physical qubit is labeled by a binary string $\boldsymbol{s}\in\{0,1\}^m$, and $|\boldsymbol{s}|$ denotes its Hamming weight, i.e., the number of nonzero entries in $\boldsymbol{s}$. 
Qubits whose labels satisfy $|\boldsymbol{s}|\geq m-r_z$ are initialized in $\ket{0}$, whereas those satisfying $|\boldsymbol{s}|\leq r_x$ are initialized in $\ket{+}$. The input state $\ket{\psi}$ is placed on the remaining $k$ qubits, whose labels satisfy $ r_x < |\boldsymbol{s}| < m-r_z$. 
The encoding is then performed using $m$ successive layers of CNOT gates. In the $\ell$th layer, CNOT gates are applied in parallel between pairs of qubits whose labels differ only in the $\ell$th bit, with the control and target directions chosen as shown in Fig.~\ref{fig:qrm_codes}(a). Repeating this procedure for $\ell=1,\ldots,m$ maps an arbitrary $k$-qubit input state to its encoded state in the QRM code space. This hypercube construction represents the simplest and most direct encoding procedure, nevertheless, it is not necessarily optimal in terms of circuit depth, gate count, or hardware connectivity, and more efficient and fault-tolerant code-specific encoding circuits may exist, for example, as presented in Ref.~\cite{menendez2024implementing}.

\subsection{Fault-tolerant implementation for 15-CNOT circuit using primal and dual $[[8,3,2]]$ code\label{app:fault_tolerant_15circuit}}
In this section, we outline a fault-tolerant implementation of the 15-CNOT circuit based on the $[[8,3,2]]$ code. 

We begin by preparing four encoded copies of the three-qubit state $\ket{\psi}$. Each copy is encoded either in the primal code, $P \equiv \operatorname{QRM}(0, 1, 3)$,  or its dual code, $D \equiv \operatorname{QRM}(1, 0, 3)$. The $i$th copy, with logical qubits $(A_i,B_i,C_i)$, is encoded into a distinct $[[8,3,2]]$ code block, denoted by $\mathsf{B}_{i}^{\mathcal{R}}$ for $i\in\{1,2,3,4\}$, with $\mathcal{R}\in\{ P, D\}$ to specify in which space, either primal, $P$, or dual $D$ the logical block lives. We denote the $j$th physical qubit of this block by $\mathsf{B}^{\mathcal{R}}_{i,j}$, where $j\in\{1,\ldots,8\}$. 
Each block is prepared using the initialization procedure described in the Appendix.~\ref{app:qrm_codes}, supplemented by the syndrome extraction, or postselection steps required by the assumed fault-tolerance model.

We initialize the first and third copies in the primal code and the second and fourth copies in the dual code. Thus, the initial code orientations are 
\begin{equation}
    \left( \mathsf{B}^P_1, \mathsf{B}^D_2, \mathsf{B}_3^P, \mathsf{B}_4^D \right) 
\end{equation} 
The first logical CNOT layer comprises two primal-to-dual CNOT layers, acting from block $1$ to block $2$ and from block $3$ to block $4$, respectively. For the first pair, the three logical CNOT gates are implemented by permuted transversal layer of eight CNOT gates. Explicitly, the CNOT layer from $\mathsf{B}_{1}^{P}$ to $\mathsf{B}_{2}^{D}$ is written as: \begin{equation}
    \overline{\operatorname{CNOT}}_{A_1\to A_2} \overline{\operatorname{CNOT}}_{B_1\to B_2}  \overline{\operatorname{CNOT}}_{C_1\to C_2} = \nonumber \bigotimes_{j = 1}^{8} \operatorname{CNOT}_{ \mathsf{B}_{1,j}^{P} \rightarrow \mathsf{B}_{2,\pi(j)}^{D}} \;. 
\end{equation}
Here, $\pi$ denotes the fixed permutation that matches the physical coordinates of the dual and primal codes. In cycle notation, $\pi=(1, 2)(4, 8)(5, 7)$ with $\pi(3)=3$ and $\pi(6)=6$.
This operation differs from an ordinary transversal CNOT between two identical primal blocks (which is described later) only through the fixed permutation of the target-qubit coordinates. Since every physical qubit participates in at most one CNOT gate, the layer remains transversal. Similarly, the CNOT layer from the third primal block $\mathsf{B}_{3,j}^{P}$ to the fourth dual block  $\mathsf{B}_{4,j}^{D}$ is implemented. 

We then apply physical Hadamard gates transversally to the first and third blocks, which are initially encoded in the primal code. As described by Eq.~\eqref{eq:qrm_hadamards}, the transversal physical Hadamard maps the primal code space to the dual code space while applying a logical Hadamard to each of the three encoded qubits. Explicitly, let $\ket{\overline{\phi}}_P$ and $\ket{\overline{\phi}}_D$ denote the primal and dual encodings, respectively, of the same logical three-qubit state $\ket{\phi}$. Then,
\begin{equation}
\left(
\bigotimes_{j=1}^{8}
H
\right)
\ket{\overline{\phi}}_{P}
= \overline H^{\otimes 3}
\ket{\overline{\phi}}_{D}. 
\label{eq:dual_to_primal_hadamard}
\end{equation}
Thus, $H^{\otimes8}$ simultaneously implements
\begin{equation}
P\longrightarrow D
\qquad\text{and}\qquad
\overline H_{A_i}\overline H_{B_i}\overline H_{C_i}.
\end{equation}
After this operation, the first and the third blocks are converted from the primal code to the dual code. Thus, all four logical blocks are now encoded in the dual code: 
\begin{equation}
        \left( \mathsf{B}^D_1, \mathsf{B}^D_2, \mathsf{B}_3^D, \mathsf{B}_4^D \right) 
\end{equation}

The remaining logical CNOTs can therefore be implemented using ordinary same-coordinate transversal CNOTs between identical dual code blocks~\cite{nelson2025small}:
\begin{equation}
\overline{\operatorname{CNOT}}_{A_1\rightarrow A_3}
\overline{\operatorname{CNOT}}_{B_1\rightarrow B_3}
\overline{\operatorname{CNOT}}_{C_1\rightarrow C_3}
=
\bigotimes_{j=1}^{8}
\operatorname{CNOT}_{
\mathsf{B}_{1,j}^{D}
\rightarrow
\mathsf{B}_{3,j}^{D}
}.
\end{equation}
Likewise, we implement three logical CNOT gates in parallel between  $(A_2, B_2, C_2)$ and $(A_4, B_4, C_4)$. 

Finally, the required non-Clifford operations are implemented by magic-state injection or gate teleportation, using the appropriate encoded resource states supplied by a magic-state factory~\cite{chang2026smallest}.

\end{document}